\documentclass{aa}

\usepackage{graphicx}
\usepackage{txfonts}
\usepackage{amssymb}
\usepackage{mathtools}
\usepackage{color}
\usepackage{siunitx}

\usepackage{footnote}
\makesavenoteenv{tabular}
\makesavenoteenv{table}
\usepackage[bottom]{footmisc}
\usepackage{tablefootnote}
\usepackage[flushleft]{threeparttable}
\usepackage{adjustbox}
\usepackage{subcaption}
\usepackage{float}

\usepackage{hyperref}

\hypersetup{
colorlinks = true,
citecolor = blue
}

\begin{document} 

\title{Flattened structures of dwarf satellites around massive host galaxies in the MATLAS low-to-moderate density fields}

\author{Nick Heesters\inst{1}, Rebecca Habas\inst{2}, Francine R. Marleau\inst{1}, Oliver M\"uller\inst{2}, Pierre-Alain Duc\inst{2}, M\'elina Poulain\inst{1}, Patrick Durrell\inst{3}, Rubén Sánchez-Janssen\inst{4}, Sanjaya Paudel\inst{5}}
   
\authorrunning{Heesters et al.}
\titlerunning{Flattened structures of dwarf satellites}

\institute{Institut f{\"u}r Astro- und Teilchenphysik, Universit{\"a}t Innsbruck, Technikerstra{\ss}e 25/8, Innsbruck, A-6020, Austria
\and
Observatoire Astronomique de Strasbourg  (ObAS), Universite de Strasbourg - CNRS, UMR 7550 Strasbourg, France
\and
Youngstown State University, One University Plaza, Youngstown, OH 44555 USA
\and
UK Astronomy Technology Centre, Royal Observatory, Blackford Hill, Edinburgh, EH9 3HJ, UK
\and
Department of Astronomy and Center for Galaxy Evolution Research, Yonsei University, Seoul 03722
}

\date{Received 26 April 2021 / Accepted 16 August 2021}

 
  \abstract
   {It was first observed in the 1970s that the dwarf galaxies surrounding our Milky Way, so-called satellites, appear to be arranged in a thin, vast plane. Similar discoveries have been made around additional galaxies in the local Universe such as Andromeda, Centaurus A, and potentially M83. In the specific cases with available kinematic data, the dwarf satellites also appear to preferentially co-orbit their massive host galaxy. Planes of satellites are rare in the lambda cold dark matter ($\Lambda$CDM) paradigm, although they may be a natural consequence of projection effects. Such a phase-space correlation, however, remains difficult to explain. In this work we analyzed the 2D spatial distribution of 2210 dwarf galaxies around early-type galaxies (ETGs) in the low-to-medium density fields of the "Mass Assembly of early-Type GaLAxies with their fine Structures" (MATLAS) survey. Under the assumption that the dwarfs are satellite members of the central massive ETG, we identified flattened structures using both a variation in the Hough transform and total least square (TLS) fitting. In 119 satellite systems, we find 31 statistically significant flattened dwarf structures using a combination of both methods with subsequent Monte Carlo (MC) simulations with random data. The vast majority of these dwarf structures lie within the estimated virial radii of the massive host. The major axes of these systems are aligned better than \ang{30} with the estimated orientation of the large-scale structure in nine (50\%) cases. Additional distance measurements and future kinematic studies will be required to confirm the planar nature of these structures and to determine if they are corotating systems.}

               
   \maketitle
%

\section{Introduction}
\label{sec:intro}
Dwarf galaxies are the lowest luminosity galaxies with the highest dark matter to stellar mass ratio in our Universe \citep[e.g.,][]{1991AJ....102..914M,2009ApJ...704.1274W,2020MNRAS.491.3496C} making them excellent candidates to study dark matter and its effects. According to lambda cold dark matter ($\Lambda$CDM), the standard model of cosmology, they are also the oldest and most numerous galaxies in the Universe, believed to be the building blocks of the higher mass galaxies we see today \citep{2012AnP...524..507F}. For those reasons they can deliver extremely valuable insight into the formation and evolution of galaxies \citep{Revaz_2018}. \par
Lambda cold dark matter cosmology predictions are successful on large scales ($\sim$\,1 -- 15000 Mpc; e.g., \citealt{white1993baryon,ostriker1995observational,eisenstein2005detection}), but a number of problems arise on smaller scales on the order below the typical spacing between galaxies ($\lesssim 1$\,Mpc; \citealt{kroupa2010local,bullock2017small}). While issues such as the missing satellite \citep{klypin1999missing,moore1999dark}, the "Too-Big-to-Fail" \citep{boylan2011too}, and the cusp-core problem \citep{flores1994observational,moore1994evidence} can increasingly be resolved by including baryonic physics and by altering the properties of dark matter \citep{simon2007kinematics,read2016dark}, the planes of satellite galaxies problem remains unsolved. \par 
\citet{kunkel1976magellanic} and \citet{lynden1976dwarf} first discovered that the dwarf galaxies and globular clusters around the Milky Way (MW) are distributed in a large circular polar ring, same as the Magellanic Stream. \citet{kroupa2005great} were the first to look at the observation in light of the $\Lambda$CDM paradigm. They found that the positions of the 11 classical MW satellites appear to align almost perpendicularly to the MW disk. Such a phenomenon should be very rare in the framework of the standard model of cosmology. \citet{kroupa2005great} assumed isotropic distributions of satellites to be a representation of the cosmological expectation, and they therefore rejected the null-hypothesis that this satellite structure was arranged from an isotropic parent distribution at random. Today, this structure is known as the vast polar structure (VPOS) of the MW, which is comprised of dwarf galaxies, globular clusters, and streams (gaseous or stellar with a globular cluster or dwarf galaxy origin). The VPOS has a root-mean-square (rms) thickness of $r_{\perp}$ =  19.9\,kpc and an rms axis ratio (shortest over longest axis) of $c/a$ = 0.209 \citep{pawlowski2012vpos,pawlowski2013dwarf,2020MNRAS.491.3042P}. \par
Currently, ten similar structures have been proposed around other galaxies \citep{libeskind2019orientation}. In addition to the VPOS, there are two planes around M31 \citep{ibata2013vast,conn2013three,shaya2013formation,2019arXiv190802298S}, one, possibly two, planes around Centaurus A \citep{tully2015two,muller2016testing,muller2018whirling,muller2019dwarf,muller2021coherent}, potentially one around M101 \citep{muller2017m,anand2018robust}, and one around M83 \citep{muller2018distances}. In addition to these satellite planes, there have been three discoveries of non-satellite planes (dwarfs located outside the virial volume of any galaxy) in the Local Group (LG): the Local Group Planes 1 and 2 (LG$_{P1}$, LG$_{P2}$; \citealt{pawlowski2013dwarf}) and the Great Northern Plane \citep[GNP;][]{pawlowski2014perseus}. \par
Phase-space correlations such as planar structures are not a frequent occurrence in cosmological simulations. In the Millennium-II (dark matter only) simulation \citep{boylan2009resolving} used in \citet{pawlowski2014co_sat}, the flatness and orbital orientation of the 11 classical MW satellites could only be reproduced in 0.3\% of realizations. The ELVIS simulations \citep{garrison2014elvis} considered the entire LG arrangement of massive galaxies and satellites, and supports the claim by \citet{pawlowski2014co_sat}, concluding that such structures occur in 0.2\% of realizations. Similar simulations have found probabilities in the range of 0.04-0.17\% \citep{pawlowski2014co_sat,ibata2014thousand} in the case of the Great Plane of Andromeda (GPoA). The Centaurus A Satellite Plane (CASP) has been studied in both the dark matter only Millennium-II simulation \citep{boylan2009resolving} and the hydrodynamical (dark matter + baryonic physics) Illustris simulation \citep{vogelsberger2014properties}, considering the on-sky flattening and the kinematic correlation along the major axis of the on-sky distribution; in the Millennium-II simulation, planes as or more extreme than the one observed are found in 0.1\% of cases while flattened, kinematically correlated structures are simultaneously identified in 0.5\% of the realization in Illustris \citep{muller2018whirling}. This demonstrates that the planes of satellites problem cannot be resolved by including baryonic physics \citep[see also][]{muller2021coherent}. \par
There are three general scenarios, outlined in \citep{pawlowski2018planes}, that are considered in attempting to explain these satellite planes in the framework of the standard model of cosmology: 1) The dwarfs were formed in the early stages of the Universe, independent from each other and have come to share a common structure and kinematics \citep{2009MNRAS.399..550L}, 2) the dwarfs were formed or assembled in some event that happened later during the evolution of the Universe \citep{2013MNRAS.431.3543H,2016ApJ...818...11S}, or 3) they represent a chance alignment and do not form stable, coherent moving structures \citep{2015ApJ...809...49B,2020ApJ...897...71S}.
No solution has been yet found that presents a satisfactory explanation of all observational evidence \citep{pawlowski2018planes}. \par
Assuming the satellite planes are coherent structures (case one and two above), four formation scenarios have been put forward to explain the frequent occurrence of these structures: \\

\noindent \textbf{Scenario 1:} \textit{Accretion along filaments of the cosmic web}. \\
Dwarfs are accreted along filaments of the cosmic web together with massive galaxies and therefore come to share an orientation in space and a common orbit \citep[][]{zentner2005anisotropic,libeskind2005distribution,lovell2011link,libeskind2011preferred}. Although this hypothesis might explain some anisotropy in the spatial distribution, \citet{libeskind2014universal} find that this effect can only enhance accretion along the direction of slowest collapse by a factor of two compared to isotropy. Additionally, it cannot explain the highly flattened structures since the scales of the cosmic filaments are much larger than galaxy scales \citep{pawlowski2013rotationally,pawlowski2012filamentary}. \\

\noindent\textbf{Scenario 2:} \textit{Group infall of dwarf galaxies}. \\
Another hypothesis suggests that satellite planes are created when a tight group of dwarf galaxies are accreted onto a massive halo. The gravitational pull of that halo would disrupt the group and form a disk-like structure with a common orbital orientation and angular momentum \citep{lynden1995ghostly}. Albeit an uncommon phenomenon \citep{wang2013spatial}, this scenario could explain some of the planar structures observed in the Local Volume \citep{li2008infall,wang2013spatial,smith2016formation}. \citet{metz2009did}, however, claim that the observed dwarf rich groups are not compact enough to explain the observed phase-space correlation in these structures. \\

\noindent\textbf{Scenario 3:} \textit{Tidal Dwarf Galaxies}. \\
Satellite planes could also be formed during interactions of massive galaxies. When two (typically late-type) galaxies come in contact with each other, tidal tails form through their mutual gravitational pull \citep{wetzstein2007dwarf}. These extended structures orbit the interaction center many times in the merging or fly-by process. In some cases, these tidal tails eventually collapse under their own gravity, forming distinct objects which themselves become hosts of star formation, the so-called tidal dwarf galaxies (TDG). TDGs may solve the satellite plane problem because multiple galaxies can form from the same tidal tail, resulting in a shared orbital direction and plane \citep{pawlowski2018planes}. However, the theory of TDG formation cannot explain the observed composition of dwarf galaxies.
Since the TDGs stem from massive galaxies, their material should show a significantly higher metallicity than classical dwarf galaxies \citep[e.g.,][]{duc1994recycled,duc1998young,weilbacher2003tidal,croxall2009chemical,reverte2007recent,kirby2013universal}. In addition, no property trends suggesting a common history have been noted in the plane members \citep{collins2015comparing}. Another problem with this hypothesis lies in the dark matter content. Since the gravitational potentials of TDGs are too shallow to carry large portions of dark matter from the original host with them, these objects are expected to be essentially dark matter free \citep{barnes1992formation,duc2004top,bournaud2006tidal,kaviraj2012tidal}. Analysis of the kinematics of multiple satellites in the LG, however, suggests mass-to-light ratios of more than $M/L = 10$\,$M_{\odot}/L_{\odot}$ \citep[e.g.,][]{2010MNRAS.406.1220W,2014ApJ...793L..14M,2018A&A...618A.122T,2020A&A...635A.152T}, meaning they have a very high dark matter content \citep{mcconnachie2012observed}. Altering gravity models could make the TDG scenario consistent with observational evidence \citep[e.g.,][]{banik2018origin,bilek2018mond}. \\

\noindent\textbf{Scenario 4:} \textit{Satellites of merging hosts}. \\
A fourth formation scenario is described in \citet{smith2016formation}, who conducted simulations of a merger between a primary and a secondary galaxy while placing a dwarf satellite population in the secondary system. During the interaction process between the two massive hosts, the satellites in the secondary system evolve into an extended, thin and rotating plane. This process is similar to the TDG formation scenario with the difference that no new dwarf galaxies are created in the process but are preserved from the premerger set-up. In this scenario the high dark matter content in the resulting satellite population does not pose a problem as it does in the previously described TDG hypothesis. Smith et al. found that the most important parameters for this event are: 1) a small extension of the dwarf population along the z-axis (perpendicular to the plane of interaction), 2) the alignment of the velocity vectors of the dwarfs with the plane of interaction and 3) the mass ratio between the merging galaxies. Satellites were added to the primary as well as the secondary system to test the influence of the mass ratio. If the mass ratio exceeds 1:2, then only satellites from the secondary system will form a corotating disk, while satellites from the primary will show a more spherical distribution. These share the disk's rotational direction with a smaller velocity. If the mass ratio is high then only a small dwarf population is expected in the secondary system compared to the primary one. This would then result in a satellite plane with few members. In cases with a close to equal mass merger dwarfs from both systems can form a flattened structure. Such a major merger, however, might also lead to the destruction of any a priori disk-like structures and suppress the formation of a new plane of satellites in this process. \citet{muller2021coherent} tested this prediction in cosmological simulations. They found, however, no correlation between merger history and the prevalence of a significant flattened, kinematically correlated structure.\par

In order to better understand the origin, nature, and prevalence of these planar dwarf structures, more satellite systems outside of the LG
need to be investigated. In this work we examine the spatial distribution of 2210 dwarfs galaxies that were identified in the "Mass Assembly of early-Type GaLAxies with their fine Structures" (MATLAS) deep optical imaging survey \citep{duc2015atlas3d}, which targeted nearby early-type galaxies (ETGs) in low-to-moderate density environments and their surroundings in 1 square degree fields. This data set presents an extraordinary opportunity to determine whether these highly flattened planar structures observed in the local Universe are a statistical outlier or a common phenomenon. \par
This paper is structured as follows: in Sect. \ref{sec:data} we describe the MATLAS data with a brief outline of the data reduction leading to the final dwarf catalog, in Sect. \ref{sec:methods} we discuss the methods used for the identification of flattened dwarf structures, while in Sect. \ref{sec:results} we present our results and we conclude our findings in Sect. \ref{sec:conclusions}.

\section{Data}
\label{sec:data}

The dwarf satellites used in this work were identified in the MATLAS deep optical imaging survey \citep{duc2015atlas3d}. MATLAS, together with the Next Generation Virgo Cluster Survey \citep[NGVS;][]{ferrarese2012next}, acquired the optical imaging of ETGs for the larger ATLAS$^{3D}$ legacy program \citep{cappellari2011atlas3d}, which aims to study the assembly and evolution of a complete volume limited sample of ETGs in the nearby Universe. The depth of the imaging, necessary to study the stellar populations in the outer regions of the ETGs and the fine structures (tidal tails, stellar streams, shells)
surrounding them, along with the relatively large field of view (FOV), are also ideal for studies of the nearby dwarfs and globular clusters \citep[][]{habas2020newly,bilek2020census}.

\subsection{MATLAS survey}
\label{subsec:matlas}

MATLAS observations were performed using MegaCam on the 3.6\,m Canada-France-Hawaii Telescope (CFHT) from 2010 until 2015 in the \emph{g}, \emph{r}, \emph{i}, and \emph{u} band using a similar strategy as its progenitors. In order to maximize detectability of low surface brightness objects and to correct for scattered light in the MegaCam, the Elixir LSB data reduction pipeline, which was designed for the NGVS project \citep{ferrarese2012next}, was used. Individual exposures were taken with offsets on the sky (2\,-\,14 arcmin; \citealt{habas2020newly}), then the images were sky-subtracted and stacked \citep{duc2012probing}. \par
The final images are $\approx 63^{'}\,\times\,69^{'}$ in size with the ETG of interest near the center. In total 150 fields were imaged by MATLAS in the \emph{g} band, down to a surface brightness limit of 28.5 - 29\,mag/arcsec$^{2}$. These fields contain a total of 180 ETGs and 59 late-type galaxies (LTGs). The resolution is $0.187^{''}$/pix, although $3\times 3$ binned images with a final resolution of $0.56^{''}$/pix were used for the subsequent identification of the dwarfs, as this promotes the detectability of faint objects \citep{habas2020newly}. 

\subsection{Dwarf catalog}
\label{subsec:dwarf_cat}

While there are different software to potentially create a fully automatic sample of dwarfs (e.g.,: \textsc{Source Extractor}; \citealt{bertin1996sextractor}, \textsc{MTObjects}; \citealt{teeninga2015improved}, or \textsc{Noise Chisel}; \citealt{2015ApJS..220....1A}), concerns about viability and completeness led the MATLAS dwarf team to apply both a visual and a semi-automatic approach. Semi-automatic dwarf catalogs are common in the literature \citep[e.g.,][]{2014ApJ...787L..37M,2018A&A...620A.165V,muller2020abundance}, and combine detection software with a visual inspection of the candidate list to remove any false detections. The final MATLAS dwarf catalog contains 2210 dwarf galaxies which have been classified as dwarf ellipticals (dEs, 73.4\,\%) and dwarf irregulars (dIrrs, 26.6\,\%). The derivation of additional properties, such as the half light radius $R_{e}$, the Sersic index $n$ and the axis ratio $b/a$ of the dwarfs, was done using the software \textsc{galfit} \citep{peng2002detailed,peng2010detailed}, a two-dimensional surface brightness fitting algorithm developed for the structure modeling of galaxies in astronomical images with an emphasis on the detailed inspection of light profiles \citep[][]{poulain2021structure}. \par

MATLAS is strictly an imaging survey, but we were able to obtain distance estimates for a number of the dwarf candidates from overlapping, complementary surveys. In total, distances for 325 dwarfs ($\sim$\,15\,\% of the final catalog) could be estimated in this way and 99\% of this subsample was confirmed to be a dwarf based on an absolute magnitude cut of M$_g$= -18 (\citep[][]{habas2020newly,poulain2021structure}. Details about the methods leading to the final dwarf catalog and its properties can be found in \citet{habas2020newly}. \par

For this work, it is crucial to determine whether or not the identified dwarfs are satellites of the massive galaxies in the images, or if their apparent proximity to the ETGs and LTGs is due to projection effects. This was tested in \citet{habas2020newly}, based on the subsample of dwarfs (325) with distance measurements obtained from other surveys. In that study it was found that 88\% of the dwarfs are satellites of the host galaxy, which was identified using the smallest 3D separation. The most likely host can only be identified this way for the dwarfs with independent distance measurements, however. Testing the reliability of the host identification by assuming that either 1) the dwarfs are associated with the targeted ETG in the field, or 2) the dwarfs are satellites of the ETG or LTG with the smallest on-sky angular separation, yields approximately the same accuracy: 64\% and 63\% of the dwarfs are satellites of the host identified in the two tests, respectively. Therefore, both assumptions lead to a similar error, namely the satellite nature of $\sim$\,64\% of the dwarfs (based on on-sky angular separation only) when the real percentage (based on smallest on-sky angular separation + radial distance) is 88\%. For simplicity and consistency, we therefore assumed that all dwarfs are associated with the targeted ETG in each field. \par

Some of the fields contain groups of massive galaxies at similar distances, whereas some fields feature multiple massive galaxies (ETG and LTG) with a wide spread in distances; these are flagged in \citet{poulain2021structure}. In these cases, the association of dwarfs with host galaxies (and therefore estimating their distances) is difficult. We initially searched for flattened structures irrespective of the number of massive galaxies in the image. However, in certain sections of the analysis, we restrict the sample to the fields (67) with a single ETG (\ref{subsec:corr}), while in a subsection of \ref{subsec:corr} ("Estimated orientation of the large-scale structure") we explicitly compare the orientation of the flattened structures against that of the massive galaxies. \par

As we are interested in the spatial distribution of the dwarfs, it is of central importance that detections are not affected by light contamination due to a bright source or the effects of cirrus emission in a part of the field. It is possible that such contamination is the reason some low axis ratio structures of dwarfs arise if on either or both sides detection is impossible. In \citet{bilek2020census} all fields were inspected for contamination by stellar halos and cirrus. We compared the number distribution of detected dwarfs in fields with and without contamination and found, however, no indication that either bright halos or cirrus emission affect dwarf detectability overall. Detailed results of this analysis are shown in Appendix~\ref{appendix:1}. 

\section{Methods}
\label{sec:methods}

To investigate the prevalence of potential flattened dwarf structures in this data set, we used two different automatic detection methods, utilizing their individual strengths. In the first subsection, we describe these methods, namely a variation in the Hough transform and total least square (TLS) fitting. The second subsection discusses a viewing angle simulation which estimates the percentage of flattened structures missed due to viewing angles which do not allow for an edge-on observation. In a third subsection we address a bias induced by the rectangular nature of the FOV in our fields.

\subsection{Flattened structure identification}
\label{subsec:flattened_str}

An initial visual inspection of the fields suggested a nonhomogeneous distribution of the dwarf galaxies. However, modeling was needed in order to eliminate potential subjectivity resulting from a purely visual examination, to test statistical significance and to quantify the properties of these structures. We therefore explored methods of automatic detection and ways to assign statistical significance to the detections, which will be described below. \par 
Previously, \citet{ibata2013vast} generated 27 (the number of satellites in M31) three-dimensional dwarf positions by randomly drawing from each dwarf's distance probability distribution function (PDF), to account for distance uncertainties. Subsequently they searched for the plane with the lowest rms thickness to subsamples of $n_{sub}$ = 15 dwarfs. This procedure was repeated 1000 times and led to a rms probability distribution of a plane consisting of 15 dwarfs. The peak value was used as a reference for further statistics. To determine the probability of a structure arranging at random given an isotropic distribution, a Monte Carlo (MC) simulation was conducted. However, concerns about the arbitrary nature of the subsample size in each field and the validity of the assumption that all dwarfs are satellites led us to explore a different approach. \par
We initially attempted to apply a simple standard least square fit, but this failed to detect structures nearing vertical orientation in the fields, because only distances to one axis are minimized. In order to look for structures in two-dimensional space with a random position angle (PA), low axis ratio $c/a$ and small scatter, the distance to the best fit line needed to be minimized in both x and y directions. This consideration led us to the Hough method and TLS fitting. As described below (\ref{HoughTransform}), the Hough method is better suited to determine the structure orientations and dimensions while the TLS method is a better fit to determine the statistical significance. \par
Neither method forces the fit to pass through the host galaxy as we were not strictly looking for satellite planes but any flattened substructures. Consequently, there are a few cases (6) in which the fit line is offset from the targeted host. 

\subsubsection{Hough transform}
\label{HoughTransform}

The Hough transform is a technique originally developed by Paul Hough \citep{hough1959machine,hough1962method}, with the goal of automatically detecting lines and other complex structures in pictures. This was done by a voting system where all points in an image would vote on a predefined set of slope and intercept pairs. Every point generates a line of possible slope and intercept pairs in the parameter space. The location where the highest number of these lines cross represents the pair with the highest number of votes at the end of this process, and this parameter pair would describe the fit that optimally describes the data points. In the ideal case of a perfectly straight line, all of the slope-intercept line pairs would intersect at a single point. For data containing scatter, however, the lines no longer define a single slope-intercept pair, although one can still obtain a reasonable approximation by identifying the region with the highest over-density of lines. We adopted a variable search area, allowing us to probe different scales and optimized it such that the structure flatness and number of voting members are maximized simultaneously. Details about this fitting technique can be found in Appendix~\ref{appendix:2}. \par
Compared to the TLS method, described in Sect. \ref{TLS-method}, the Hough transform does not rely on the quality of a first guess fit but takes the whole picture into account without making any assumptions or initially removing outliers. After the voting process is concluded and the optimal fit parameters are found, the dwarf structure is defined by the objects which voted for this best-fit line. All other dwarfs in the field are considered outliers. This method removed $\sim$\,22\% of the dwarfs, on average, or $\sim$\,4 dwarfs per field. The distributions of removed outliers in percentile and absolute values as well as a scatter plot comparing the two are shown in Appendix~\ref{appendix:3} \par
This method, however, is too computationally expensive to use in MC simulations with 10$^5$ realizations per field. While the Hough method is more suitable to determine the flattened structure orientations and dimensions, only the TLS + outlier removal method can be used to assign statistical significance to the structures. \par

\subsubsection{Total least square fitting (TLS) or orthogonal distance regression (ODR)}
\label{TLS-method}

The method of TLS fitting, or orthogonal distance regression (ODR), minimizes the distances perpendicular to the fit line as opposed to vertically, as in the standard least square fitting. The python package "scipy.odr" in combination with a linear model was used to fit the data. This method considers all objects in the field equally. However, not all dwarfs in a given field are guaranteed to be satellites of the same massive galaxy; some fields contain multiple massive galaxies, there may be foreground or background hosts, or other massive galaxies may lie just outside the FOV. Therefore it is reasonable to assume that a portion of the dwarfs in each field might not be part of any apparently flattened substructure. \par 
This method was only used as a follow-up to the Hough method. We fit all dwarfs in the field using the TLS method, and the resulting slope, intercept, and residual variance were recorded as an initial guess for the following steps. The dwarf with the largest perpendicular offset to the best fit line was then removed and the fit was recalculated using the new subsample of dwarfs. This step was repeated in a recursive loop until we had removed the same number of outliers as non-Hough-voters. \par
To determine the statistical significance of these structures after outlier removal, an MC simulation was run with random data. For every field, a sample the same size as the total number of dwarfs in the field was generated by leaving the radial distribution of the dwarfs around the central ETG unchanged and assigning a random angular position to each object. The newly generated random data was then treated in the same way as the real data. For every field the same number of outliers were removed in the data and the random samples. To increase statistical significance, for every one of the fields, $10^{5}$ random data samples were generated and analyzed as described above. The p-value of the observed flattened structure was calculated based on the number of simulated runs with residual variance values $\leq$ the residual variance of the original data, divided by the number of runs. This p-value was then used to differentiate between detection or nondetection. \par
Because the TLS method heavily depends on the initial fit where all dwarfs are included, and the fact that dwarfs at the edges of the field -- sometimes clearly not part of a clustered structure near the center -- are given equal weight in the calculation, the initial fit may lead to a best fit line which does not agree with the fit line found via the Hough method. \par
Given the specific advantages of both the TLS + outlier removal routine and the Hough transform, we analyzed all fields using both methods. Subsequently we compared the position angles of the fits resulting from the methods and noted them as agreeable if the difference is less than \ang{30}. Such a difference can result from the inclusion or rejection of a single dwarf, although the overall dwarf set remains stable. In this way, we conclude that the methods agree in 122/150 ($\sim$\,81\%) cases. In the subsequent analysis we only use cases in which the methods agree.

\subsection{Disk viewing angle rotation simulations}
\label{disk_rot_sim}

Without distance measurements, the only situation in which we as an observer would be able to detect any flattened structure is to view it nearly edge-on. We therefore evaluated the likelihood of viewing a thin disk-like system edge-on. We simulated and plotted 1000 points arranged into a disk the size of the GPoA (diameter $\sim 400$\,kpc ; perpendicular scatter $\sim 14$\,kpc) observed from different viewing angles. This can be visualized by placing the disk inside a sphere and aligning it with the equator. The observer is then randomly placed on a location on the surface of the sphere viewing the disk inside from a particular angle which leads to an apparent minor axis. In $10^{5}$ realizations, we assigned the observer random angles from the intervals $\theta_{d} \in [-\pi/2,\pi/2]$ and $\phi_{d} \in [0,2\pi)$. This resulted in apparent minor axes between 14 and 400\,kpc. \par
\begin{figure}
\centering
\includegraphics[width=\linewidth]{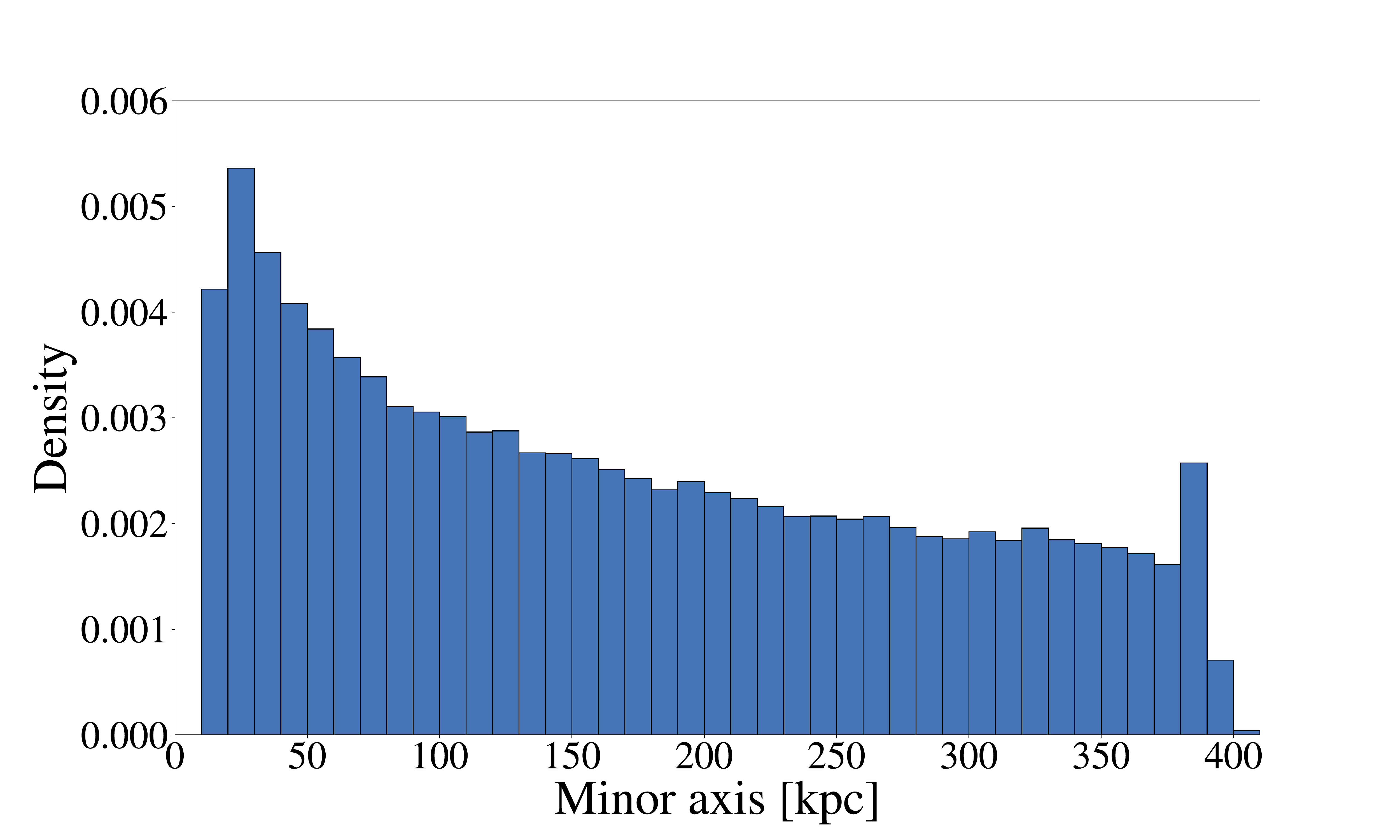}
\caption{Apparent thickness distribution from the disk rotation simulation.}
\label{figure:disk_rot}
\end{figure}

 
A histogram of the thicknesses, shown in Fig. \ref{figure:disk_rot}, demonstrates that it is more likely for the structure to appear edge-on rather than face-on. Based on our simulations, the observation of an apparent minor axis of $\approx$10\,-\,75\,kpc is most likely. There is a gentle, steady decline in realizations toward higher minor axis dimensions. \par
These simulations allow us to estimate the percentage of planes we likely missed due to our viewing angle. From our visual identifications, we can define the maximum on-sky flatness that we measure. Our estimate of the percentage of planes missed due to the viewing angle is obtained by summing up the frequencies of viewed axis ratios above that value and dividing by the total number of realizations ($10^{5}$). We therefore estimate that $\sim$\,38\,\% of flattened structures are missed due to suboptimal viewing angle, if these structures are distributed homogeneously across the Universe.

\subsection{Field of view considerations}
\label{subsec:FOV}

Due to the fact that all dwarfs in our rectangular fields are considered, best-fit lines are preferentially diagonal and therefore more extended in these cases. We addressed this bias by only including dwarfs within circular areas around the host. This leads to the exclusion of many dwarfs near the corners of our fields and thus a significant reduction in the number of statistically significant flattened structures, from 31 to 24. We report the numbers considering this bias in addition to the ones using the full FOV that is considering all dwarfs in our fields. The numbers resulting from a restriction to a circular FOV can be found in Appendix~\ref{appendix:4}.

\section{Results}
\label{sec:results}

In this section we discuss our findings. We first modeled all fields with the Hough transform and TLS + outliers method. In some cases with multiple ETGs in the field, these galaxies were individually targeted again in other fields. After excluding duplicate fields, that is with a significant overlap with others, 119 cases where the two methods agree remain. We then examined the statistical significance of the structures in this subsample. In fields featuring a statistically significant structure we determined their projected dimensions and compared the distributions to those of the well studied planes in the Local Volume. In order to assess the number of potential satellite planes in our sample, we estimated the virial radius of the host galaxies and examined the number of structures residing within. Finally we investigated correlations with the host properties including the photometric and kinematic position angles of the targeted ETG in each field, the estimated orientation of the large-scale structure, the mass of isolated host ETGs, the host ETG's rotator class, and features related to merger history. We also examined the flattened structures for evidence of corotation.

\subsection{Statistical significance of the flattened structures}

\begin{figure}
\centering
\includegraphics[width=\linewidth]{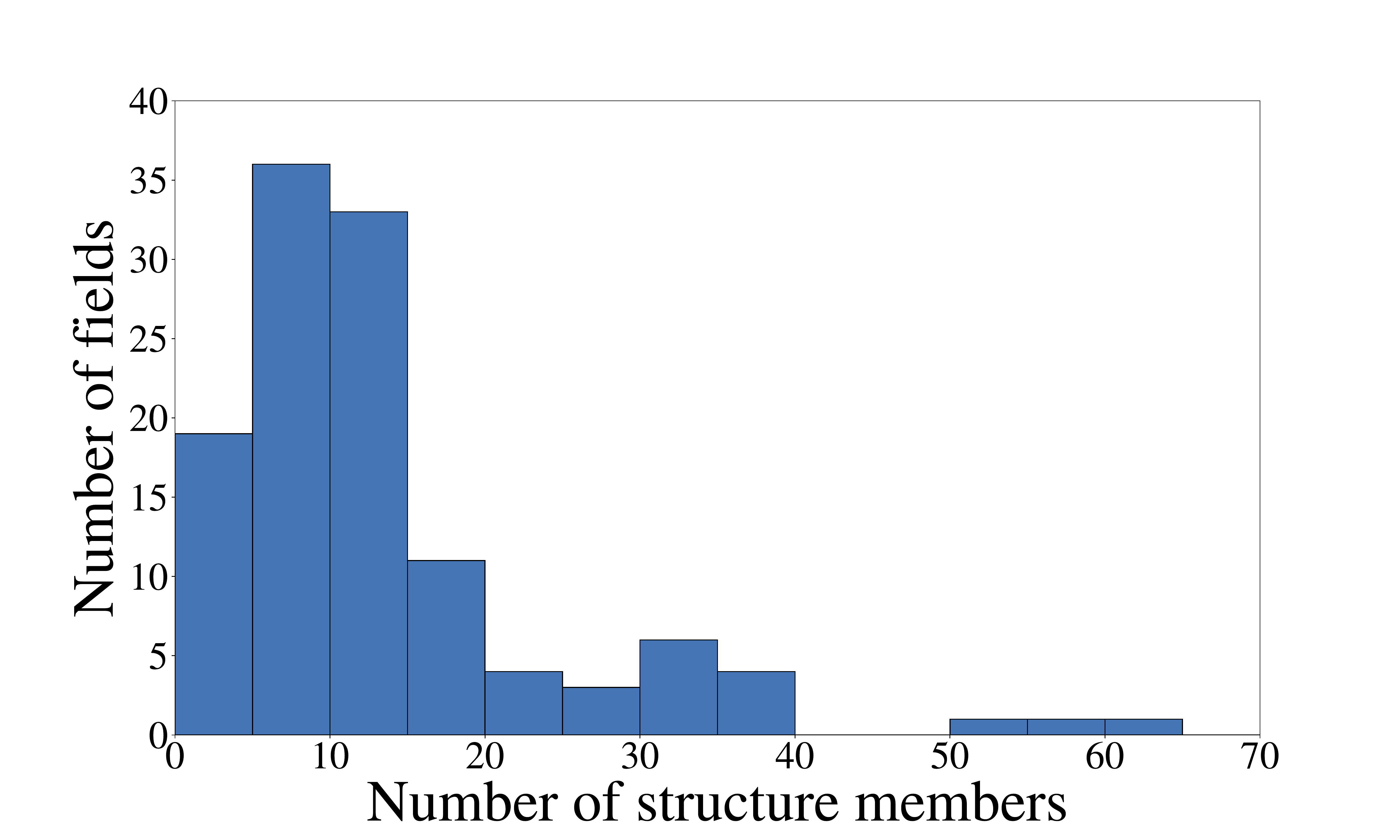}
\caption{Dwarf number distribution of Hough voters.}
\label{figure:num_dist}
\end{figure}

\begin{figure}
\centering
\includegraphics[width=\linewidth]{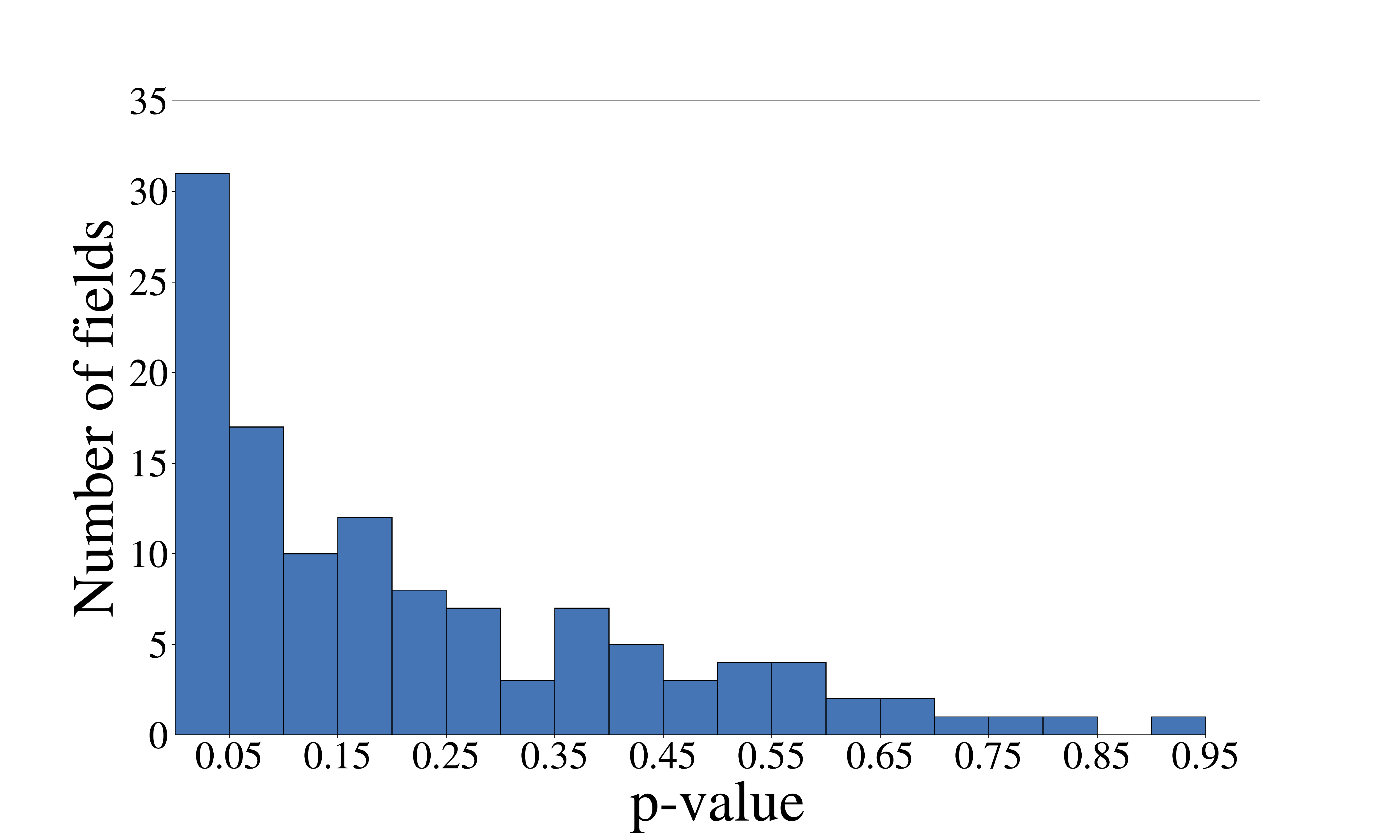}
\caption{P-value distribution of the dwarf structures in 119 fields. Only unique fields (significantly overlapping fields excluded) where the two detection methods agree were included in this analysis.}
\label{figure:pval}
\end{figure}

We used the TLS method with recursive outlier removal in an MC simulation (see Sect. \ref{TLS-method}) to determine the statistical significance of the flattened dwarf structures. Duplicate fields, as well as those cases where the TLS and Hough methods do not agree, were excluded from this analysis. In Fig. \ref{figure:num_dist} we show the number distribution of the dwarfs which voted for the same slope and intercept pair in the Hough technique (Hough voters), that is the distribution of flattened structure members. We find a median value of ten members per structure. The results of the MC simulation are shown in Fig. \ref{figure:pval}. Out of these 119 fields, 31 ($\sim$\,26\%) show a p-value p $\leq$ 0.05. In these cases we reject the null hypothesis that these structures were arranged at random assuming an isotropic parent distribution at a significance level of $\alpha$ = 0.05. Our previous estimations discussed in Sect. \ref{disk_rot_sim} suggest that 38\% of flattened structures are missed due to suboptimal viewing angles. Consequently, we estimate the detection of 31/0.62 = 50 structures given a full 3D picture of the systems. This corresponds to 42\% of fields for this study. Interestingly, \citet{ibata2014velocity} estimate that > 60\% of satellites should reside in planes, using the Millenium-II simulation \citep{boylan2009resolving}. This percentage appears comparable to our estimation. It should, however, be noted that the study by \citet{ibata2014velocity} was based on the velocity anticorrelation of diametrically opposed galaxy satellites around their host.

\subsection{Flattened structure dimensions}
\label{struc_dim}

We used the Hough fitting method to determine the best fit for the flattened structures. The subset of dwarfs which voted for the best fit parameters was used to calculate the centroid $\mathbf{r}_{0}$ of this structure as

\begin{equation}
\mathbf{r}_{0} = \frac{1}{N_{H}} \sum_{i=1}^{N_{H}} \mathbf{r}_{i}.
\label{centroid}
\end{equation}

Here $N_{H}$ is the number of dwarfs fitted by the Hough technique (Hough voters) and $\mathbf{r}_{i}$ are their position vectors. This centroid and the fit line were then used as reference to calculate the rms length and height considering only the Hough voters. However, it could be argued that none of the dwarfs should be removed from the fields, as we have only a statistical motivation at present to label them as outliers. The dimensions of the planar structures, if we do not remove any of the dwarfs, are presented in Appendix~\ref{appendix:5} for comparison. \par

\begin{figure}[htb]
\centering
\begin{subfigure}[t]{\linewidth}
   \includegraphics[width=\linewidth]{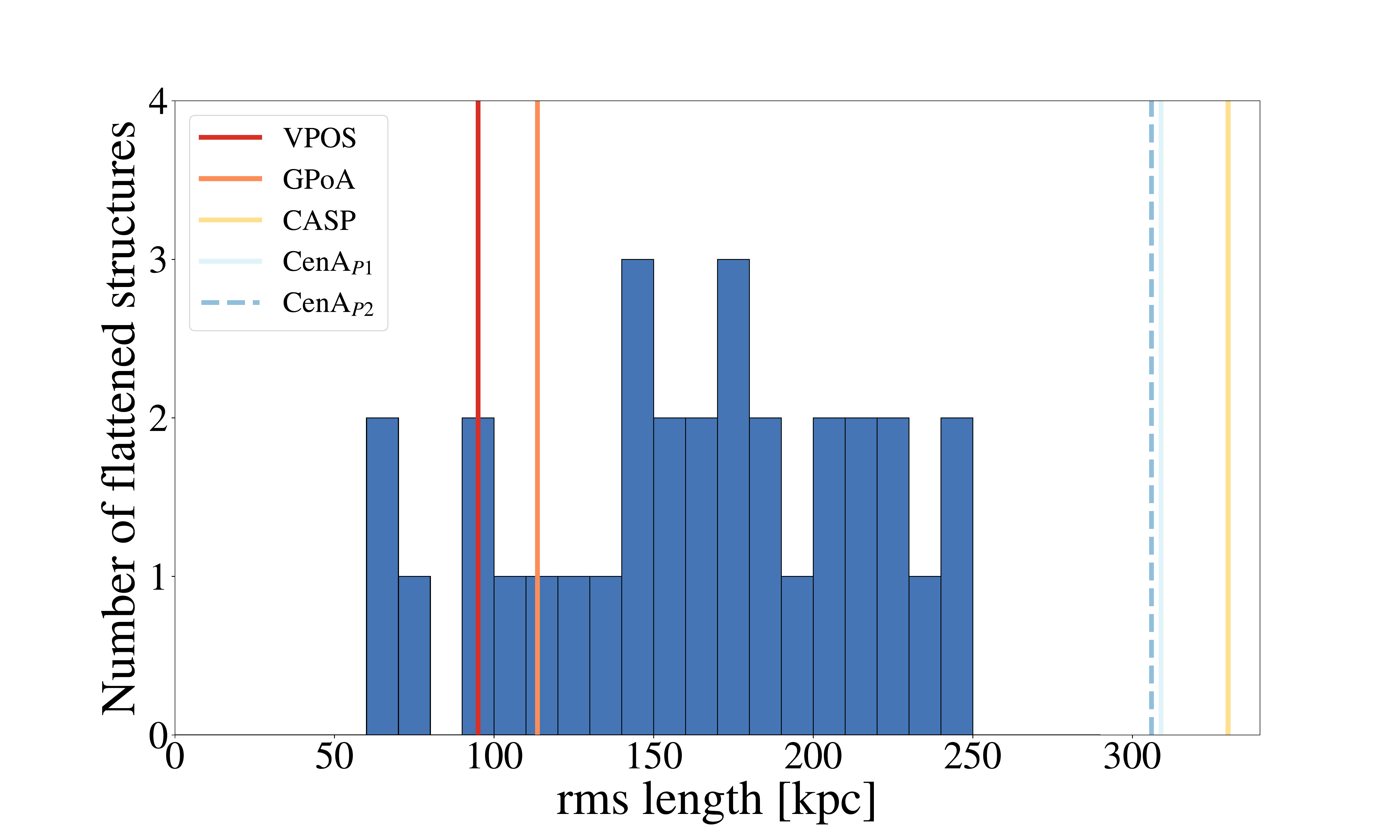}
   \caption{}
   \label{figure:dim1} 
\end{subfigure}
\begin{subfigure}[t]{\linewidth}
   \includegraphics[width=\linewidth]{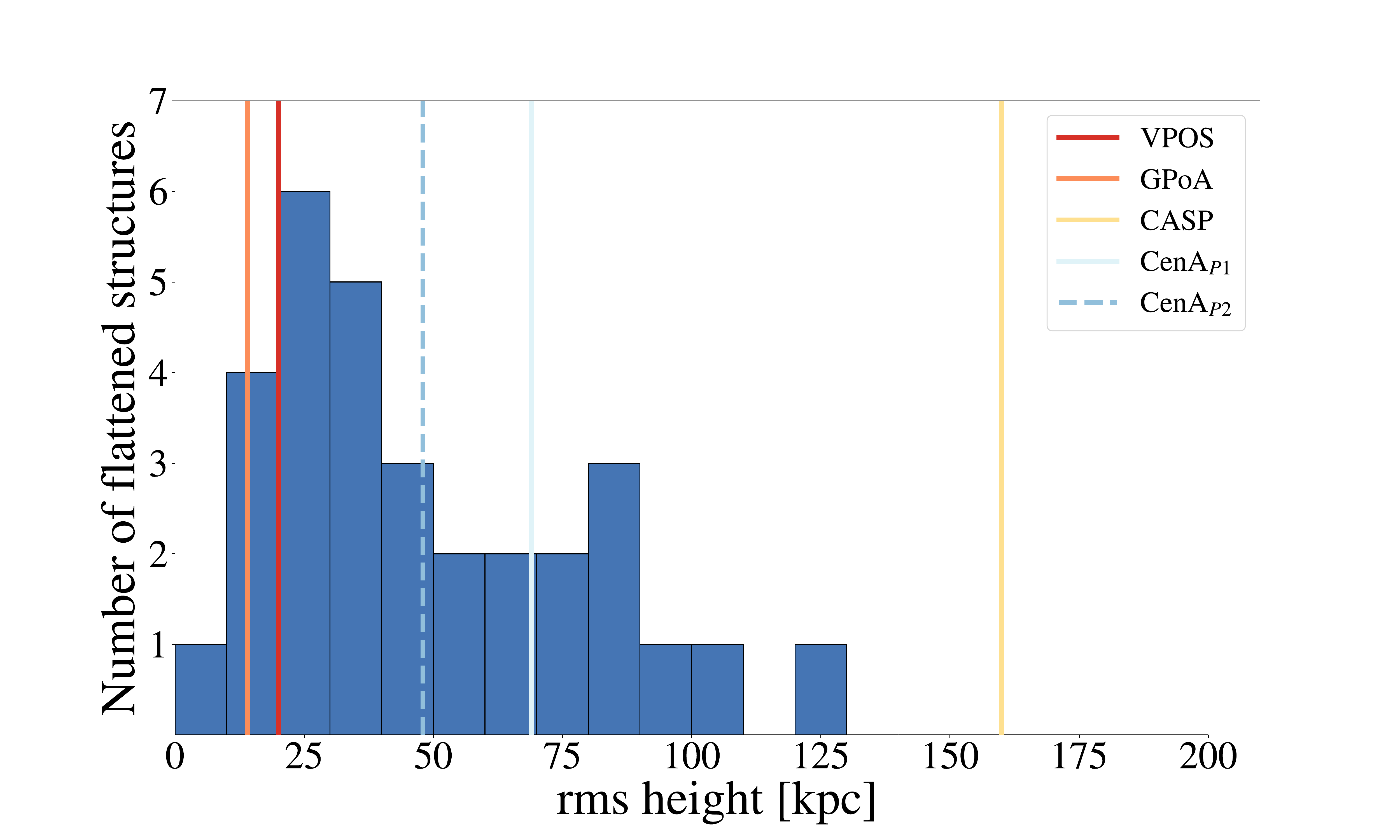}
   \caption{}
   \label{figure:dim2}
\end{subfigure}
\begin{subfigure}[t]{\linewidth}
   \includegraphics[width=\linewidth]{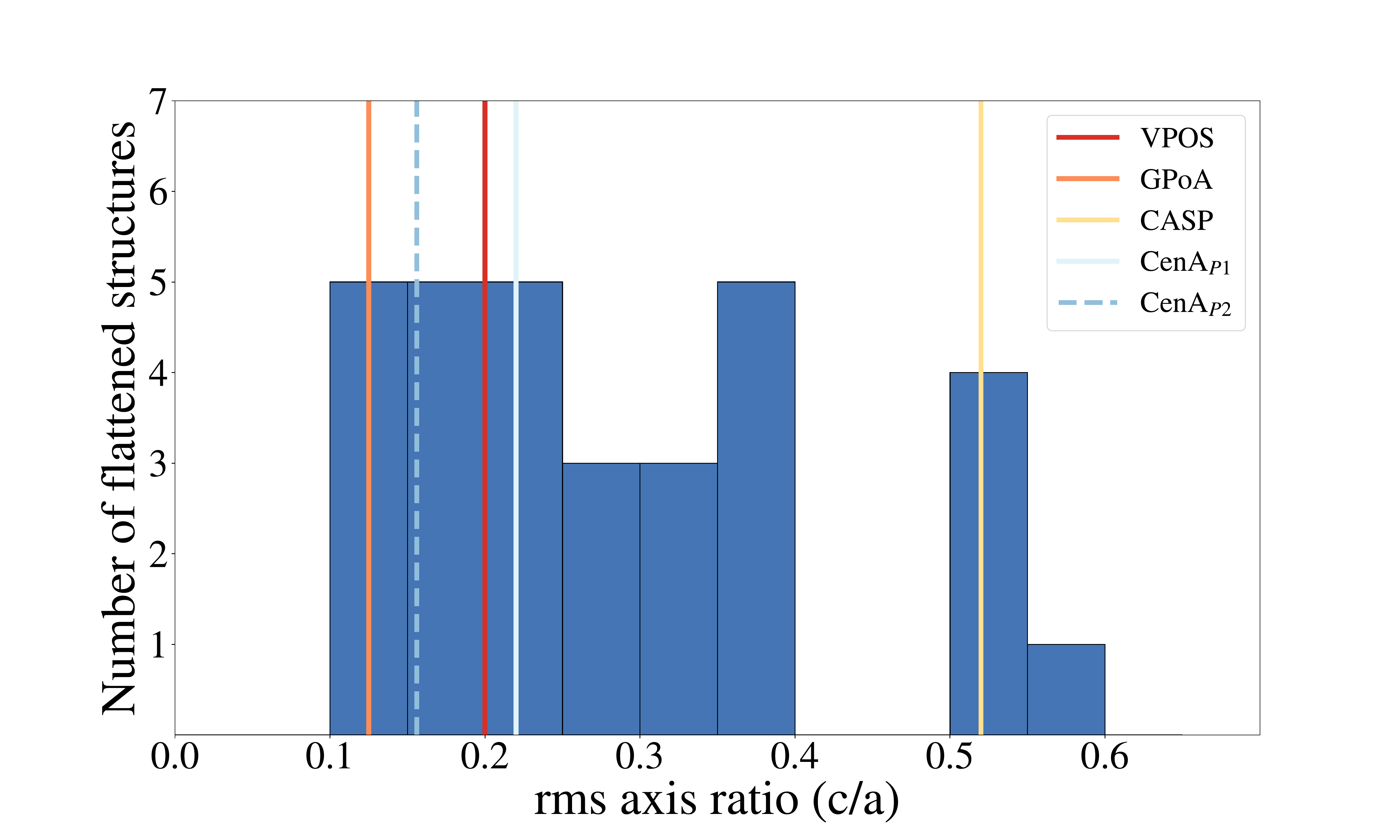}
   \caption{}
   \label{figure:dim3}
\end{subfigure}

\caption[]{Rms dimensions of the automatically detected (p $\leq$ 0.05) structures. These dimensions were calculated via the small angle approximation assuming all dwarfs are at the distance of the targeted host galaxy. Shown are the structure length (a), height (b), and axis ratio (c). Measured values of several well-studied planes are over plotted for comparison purposes. References for these are: \citet{pawlowski2013dwarf} (VPOS, GPoA), \citet{muller2018whirling} (CASP), \citet{tully2015two,muller2016testing} ($CenA_{P1}$, $CenA_{P2}$)}
\label{figure:dimensions}
\end{figure}

The distribution of the rms length ($a$) is shown in Fig. \ref{figure:dim1}. We find a relatively flat distribution with a peak at $\sim$\,150\,kpc. The values for the best studied planar structures, namely the VPOS, the GPoA \citep{pawlowski2013dwarf}, the two planes (CenA$_{P1}$, CenA$_{P2}$) in the Centaurus A system \citep{tully2015two}, and the CASP \citep{muller2018whirling,muller2019dwarf,muller2021coherent}, were added to the plot for comparison. \par
The distribution of the rms height ($c$) is shown in Fig. \ref{figure:dim2}. The distribution shows a peak at $\sim$\,25\,kpc. Based on the rms length and height distributions of the structures we detect, the majority of our detected flattened structures have properties between those of the VPOS or GPoA and the CASP. The axis ratio is then defined as rms height divided by rms length ($c/a$). A histogram for this parameter is shown in Fig. \ref{figure:dim3}. This distribution is flat with an equal number of fields in the range $\sim$\,0.1 to 0.25 and in the 0.35\,-\,0.4 bin. In this case the majority of the detected structures appear to agree well with the best studied objects in the Local Volume. As discussed in Sect. \ref{subsec:dwarf_cat}, all physical distances were calculated by using the distance of the central ETG (provided by the ATLAS$^{3D}$ team) in the field, assuming that all dwarfs in the field are at that same distance. The on-sky separations in pixels were then transformed to kpc by using the small angle approximation. \par
\begin{figure}
\centering
\includegraphics[width=\linewidth]{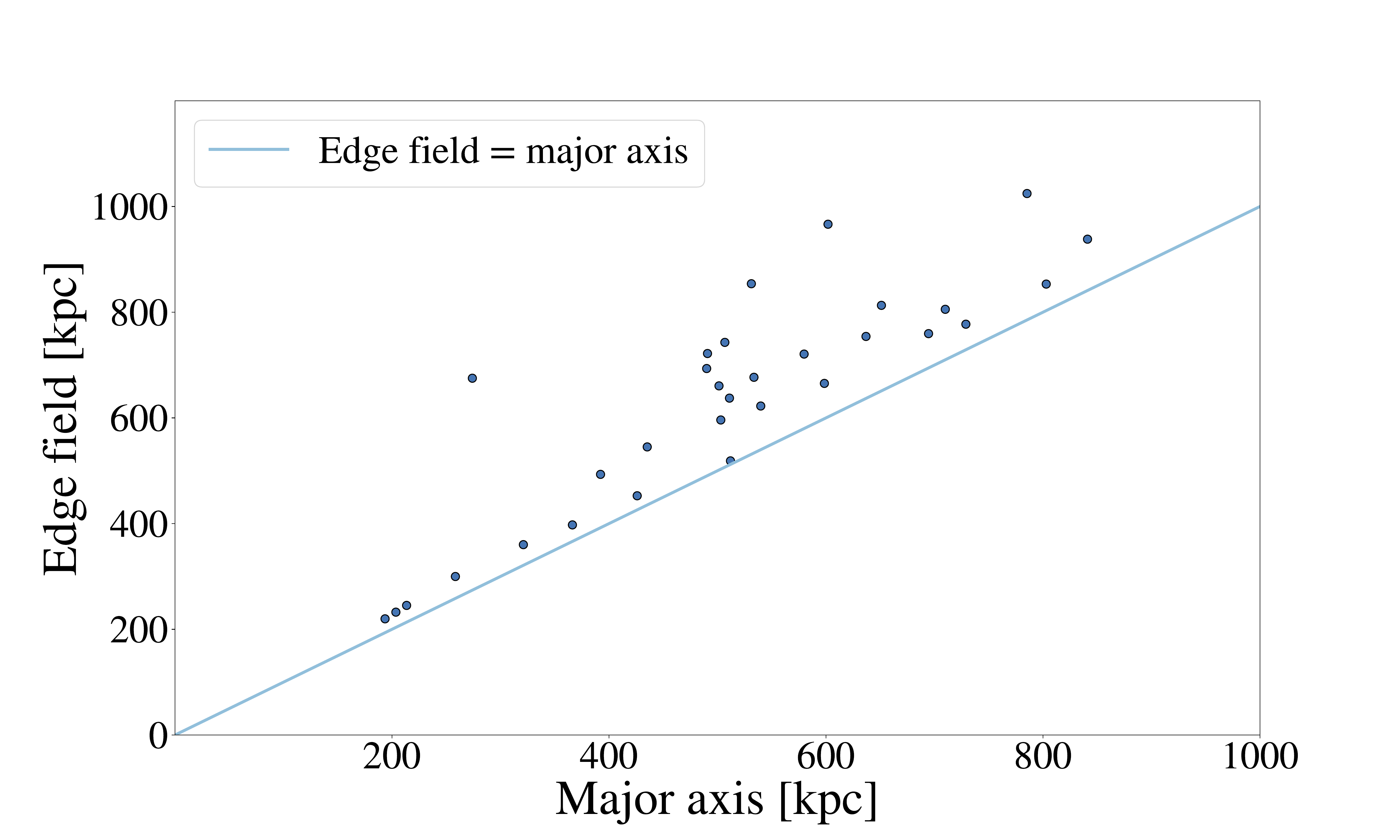}
\caption{Major axis vs field size. Testing the relationship between the long axis of the structures and the field size. The solid line shows the one-to-one correspondence.}
\label{figure:major_axis_fsize}
\end{figure}

Given the size of the fields, it is possible that the major axes that we measure have been truncated by the FOV of the MATLAS images. To test if we have detected the full extent of the structure or if it has been cut off, we plotted the major axis of the structures vs. the length of the fit line ranging the full size of the field in Fig. \ref{figure:major_axis_fsize}. The major axis was calculated by inspecting the points on the fit line corresponding to the foot of the perpendicular from the dwarf positions. Then the length of the fit line between the first and last point on the line yields the major axis of the structure. The near-linear relationship apparent in Fig. \ref{figure:major_axis_fsize} suggests that the major axes of these structures are likely constrained by the field size and should, therefore, be considered lower limits only. This in turn makes the axis ratios discussed above upper limits. \par
We further calculated the virial radius of the host galaxies, to estimate how many of the dwarfs are (likely) gravitationally bound to the system. To achieve this, we used the stellar masses calculated from the $K_{s}$ magnitudes of the massive galaxies and followed the procedure described in \citet{cui2021origin} to transform them into halo masses. These halo masses were then used as estimates of the virial masses. In order to calculate the virial radii we used the formula

\begin{equation}
    r_{vir} = \left(\frac{M \cdot G}{100 \cdot H_{0}^2}\right)^{1/3}
    \label{rvir}
\end{equation}

with $r_{vir}$ the virial radius, $M$ the galaxy mass ($M_{halo}$), $G$ the gravitational constant, and $H_{0}$ the Hubble parameter (70\,km/s/Mpc). This separation of satellite and non-satellite dwarf galaxies using the virial radius of the closest host has previously been used in defining planar structures in the local Universe, although not all of the ten thus far identified dwarf planes were strictly categorized in this way. Some cases are comprised entirely of satellite (VPOS, GPoA) or non-satellite galaxies (LG$_{P1}$, LG$_{P2}$, GNP) while others, such as the structure around M101, CenA$_{P1}$, and CenA$_{P2}$, contain dwarfs both inside and outside of the hosts virial radius \citep{libeskind2019orientation}. Of the 2210 dwarf galaxies in the sample, 2063 dwarfs fall within the virial radius of at least one massive galaxy (ETG or LTG). Consequently, for all flattened structures the majority of the dwarf members lie within the estimated projected virial radius.

\subsection{Correlation with massive host properties}
\label{subsec:corr}
In this section we investigate possible correlations between host galaxy and flattened structure properties. We conducted the following analyses in fields featuring a single massive host, since these are the systems where the association of the satellites to the host is most robust. \par

\begin{figure}
\centering
\begin{subfigure}[t]{0.7\linewidth}
   \includegraphics[width=\linewidth]{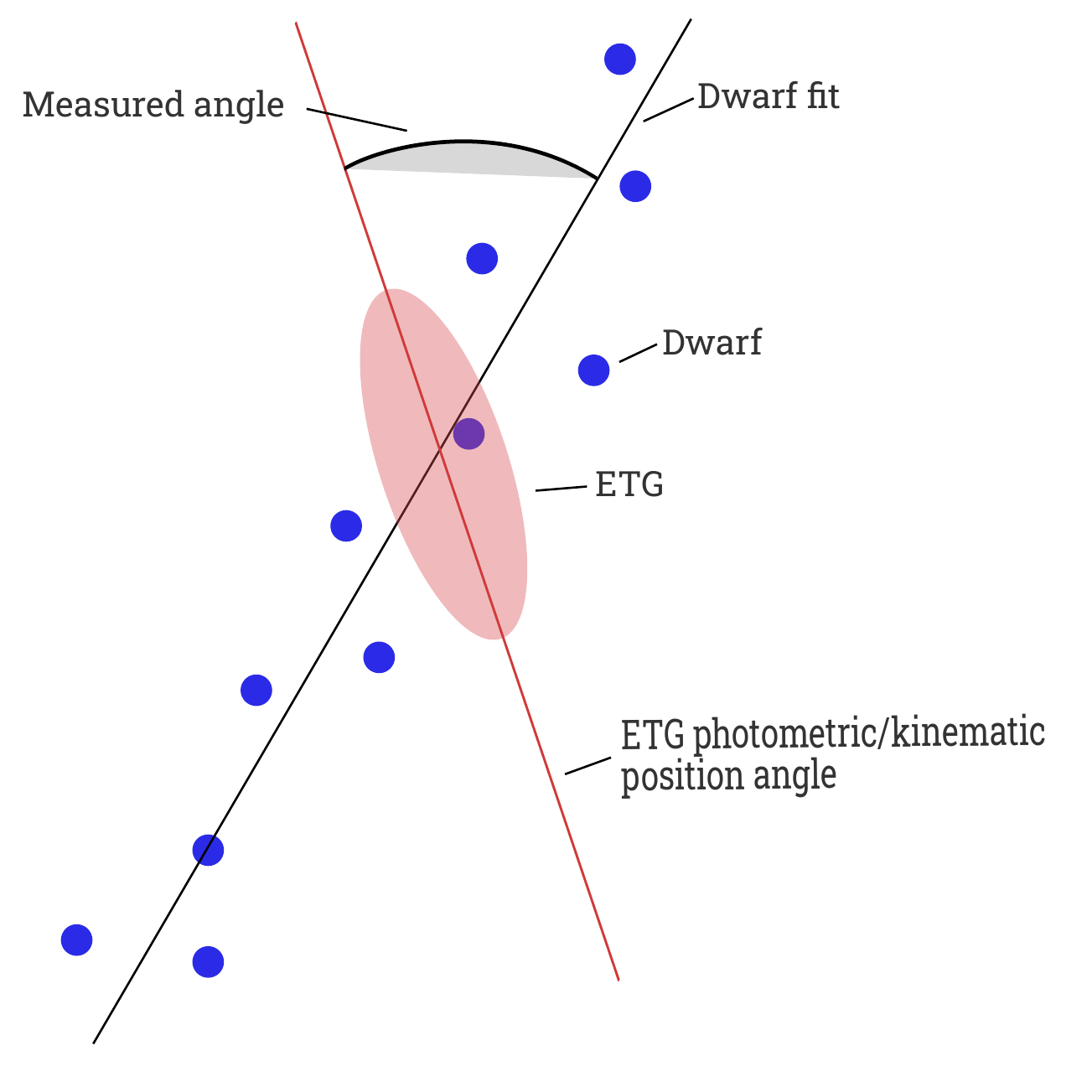}
   \caption{}
   \label{figure:ang1} 
\end{subfigure}
\begin{subfigure}[t]{0.7\linewidth}
   \includegraphics[width=\linewidth]{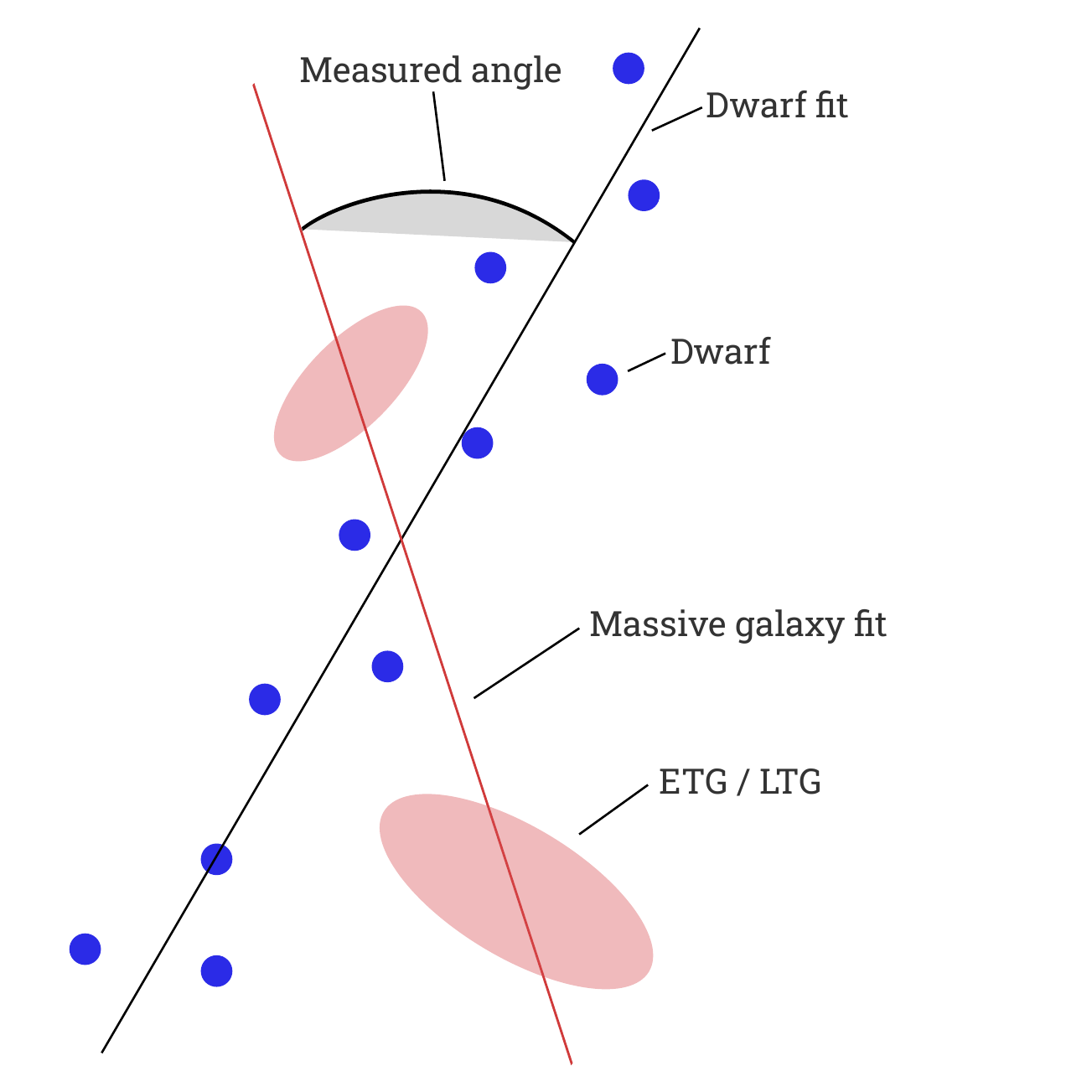}
   \caption{}
   \label{figure:ang2}
\end{subfigure}

\caption[]{(a) Illustration of the measured angles for the analysis of alignment of the flattened dwarf structure with the massive host PAs. (b) Measured angles for the analysis of dwarf structure alignment with the estimated large-scale structure. The massive galaxy fit was used as a proxy for the orientation of the large-scale structure.}
\label{figure:angles}
\end{figure}

\vspace{0.2cm}
\noindent \textit{Photometric and kinematic position angle (PA)} \vspace{0.2cm} \\
\noindent The major axes of two of the three best studied observed planar structures in the local Universe, VPOS and CASP, are almost perpendicular do the disk and dust lane, respectively. In the case of the GPoA, the angle between the plane and the M31-disk is $\sim $\,$\ang{50}$. We tested if there is evidence for such a correlation in our data set. Using the photometric and kinematic PAs of the ETGs in the field from \citet{krajnovic2011atlas3d}, we compared with the PA of the major axis of the detected dwarf structure. The photometric PA is defined as the angle between the north (top of the image) and the major axis of the structure toward east (left in the images). The kinematic PA describes the rotation of the galaxy and is defined as the angle between north and the mean receding part of stars in the velocity map of the massive host. The corresponding rotation axis of the galaxy is perpendicular to the kinematic PA \citep{krajnovic2011atlas3d}. Figure \ref{figure:ang1} shows a sketch illustrating the measured angles. \par 
When considering the photometric PA, five structures are aligned better than 30$^{\circ}$, six structures show a difference in PAs between \ang{30} and \ang{60} and two cases show differences in the range \ang{60} to \ang{90}. We therefore conclude there is no preferred alignment with respect to the photometric PA of the massive hosts. When considering the kinematic PA, five show an angle $\leq$ 30$^{\circ}$, six are in between \ang{30} and \ang{60} and two are located in the \ang{60} to \ang{90} bin. This distribution is not unexpected, as the photometric and kinematic PA are very similar for most galaxies. Similar to the photometric PA, we conclude that there is no evidence for a preferred orientation in relation to the kinematic PA of the host galaxy. Due to low number statistics for the well studied planar structures in the Local Volume, we would need more confident cases with accurate distance and kinematic data to make a definitive statement about such a correlation. \par

\vspace{0.2cm}
\noindent \textit{Estimated orientation of the large-scale structure} \vspace{0.2cm} \\
\noindent A recent study by \citet{libeskind2019orientation} suggested that the planar structures in the Local Volume align with the $e_{1}$ vector of the large-scale structure, namely the direction of the fastest collapse and close to the direction of the local void expansion. All but one (LG$_{P2}$) of the ten planar structures show an angular difference $\leq$ 60$^\circ$ to $e_{1}$. The statistics on the other two eigenvectors of the local filament are as follows: 7/10 and 8/10 of the structures are averted by $\geq$ 60$^\circ$ from $e_{2}$ and $e_{3}$ (slowest collapse), respectively. \par
\begin{figure}
\centering
\includegraphics[width=\linewidth]{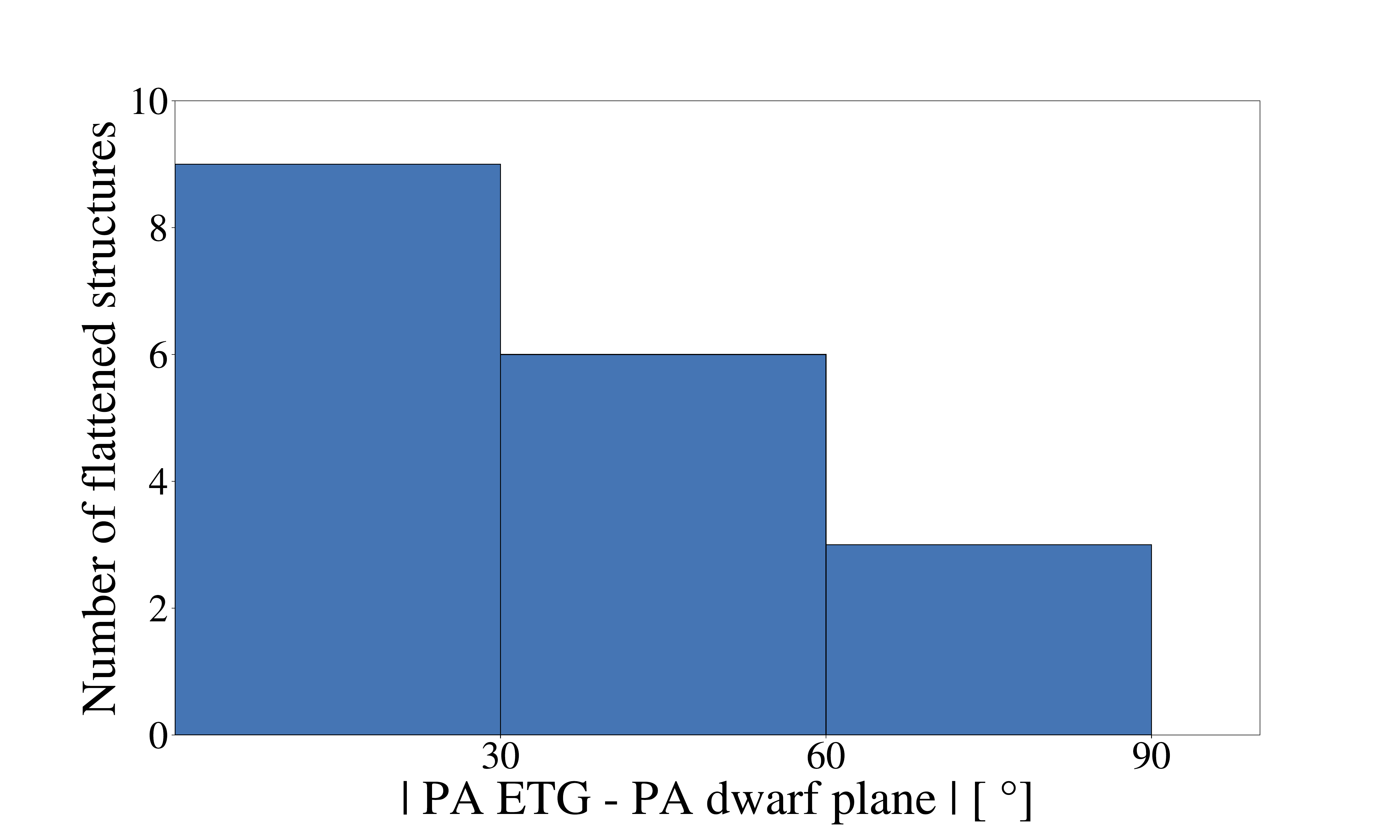}
\caption{Distribution of angles between the dwarf structure orientation and a linear fit considering the massive galaxies as a proxy to the larger scale structure.}
\label{figure:pa_large_scale}
\end{figure}

Following up on these observations, we looked for a potential correlation between the flattened structure orientation and the PA of a fit considering the massive galaxies in the field. We used the orientation of the "filament" of the massive galaxies in the field as a proxy for the orientation of the large-scale structure. A sketch visualizing the measured angle for this analysis is shown in Fig. \ref{figure:ang2}. The results for this measurement are shown in Fig. \ref{figure:pa_large_scale}. We considered the 18 fields with automatic identifications (p $\leq$ 0.05) and multiple massive galaxies and find that nine (50\,\%) structures are aligned better than 30$^{\circ}$ with the orientation of the massive hosts. In six cases the difference is between \ang{30} and \ang{60} while three structures show angles in the range \ang{60}-\,\ang{90}. \par
Due to the 2D projection and the low number statistics, claims of any correlation with the large-scale structure orientation are difficult to make. An alignment of the fit line considering the massive galaxies in the field with the major axis of the structure would be equivalent to the predicted avoidance of the $e_{3}$ vector. We cannot, however, make any statement about the orientation along the $e_{1}$ or $e_{2}$ eigenvectors. Following the stated approximations and simplifications, 50\,\% of the structures' normal vectors avoid the massive galaxy orientation by $\geq$ 60$^\circ$. We therefore find the structures are preferably aligned with the estimated large-scale structure orientation, consistent with the results from \citet{libeskind2019orientation}.

\vspace{0.2cm}
\noindent \textit{Halo mass of isolated ETGs} \vspace{0.2cm} \\
\noindent We compared the halo mass ($M_{halo}$) of isolated host ETGs with the prevalence of flattened structures. The masses in this sample range from 1.29\,$\times$\,10$^{12}$\,M$_{\odot}$ to 3.34\,$\times$\,10$^{13}$\,M$_{\odot}$. The lowest and highest mass host with a significant flattened structure have masses of 1.66\,$\times$\,10$^{12}$\,M$_{\odot}$ and 2.71\,$\times$\,10$^{13}$ \,M$_{\odot}$, respectively. We find no significant difference between the ETG mass distributions of cases with and without a flattened structure in the field. A two-sample Kolmogorov–Smirnov (KS) test yields a p-value of $p_{mass}$ = 0.23. Therefore, we cannot reject the null hypothesis that both samples were drawn from the same parent distribution and note no correlation between halo mass of the isolated host ETG and its dwarf satellite arrangement.

\vspace{0.2cm}
\noindent \textit{Kinematics of the targeted ETG in the field} \vspace{0.2cm} \\
\noindent In terms of kinematics, ETGs are generally divided into slow and fast rotators. Recent studies \citep[][]{duc:hal-02382682,bilek2020census} show that slow rotators have a higher incidence of tidal features which is thought to be related to a strong merger history. Galaxy mergers are frequently discussed as one of the possible formation scenarios for the flattened structures we observe in the local Universe (see Sect. \ref{sec:intro}). This leads us to expect an excess of such structures around slow rotating ETGs. We compared the central ETGs kinematic class \citep{emsellem2011atlas3d} with the prevalence of a flattened structure. There are ten fields where the structure detection methods agree and which feature a slow rotator as the targeted ETG. We find that four of these shows a significant flattened structure. In contrast there are 56 cases (structure detection methods agree) with fast rotators as the central ETG. In nine of these cases we detect a significant flattened structure. We therefore report that 40\% of the slow rotators and 16\% of the fast rotators in our sample host significant flattened structures. We note some excess of such structures around slow rotators, albeit not very statistically significant given the low number of slow rotators in our data.

\vspace{0.2cm}
\noindent \textit{Features related to merger history} \vspace{0.2cm} \\
\noindent Tidal tails, streams, and shells are additional tracers of a past merger histories. There are correlations between such features and a planar structure in two cases in the local Universe: the LG$_{P1}$ extends between the MW and M31 along the direction of the Magellanic stream \citep{pawlowski2018planes} and the tidal feature around Centaurus A is aligned with the CASP. Such features were recently identified by \citet{bilek2020census} in the MATLAS fields. The alignment of these structures and the flattened satellite distributions will be the subject of a future paper. Here, we compared the prevalence of these features and flattened structures. A relatively small fraction of our fields features tidal tails, streams, and shells: 17\,\%, 15.8\,\%, 12.5\,\%, respectively. We find no significant correlation between tidal tails, streams, shells and the occurrence of a flattened structure. We find seven flattened structures and 15 random configurations in fields featuring these signs of a past merger. A larger fraction of fields with such features is required to confidently test these predictions.

\vspace{0.2cm}
\noindent \textit{Rotating structures} \vspace{0.2cm} \\
\noindent Distance measurements from other surveys are available for a portion of the dwarfs in this sample. By checking the recessional velocities of these dwarfs and doing a comparison against the value of the central ETG in the field, we can search for potential rotational trends --- that is, dwarfs with relative velocities in opposite directions on opposing sides of the host --- in these structures. There are seven significant flattened structures with a single ETG in the field and with velocity measurements available on either side of the host. We find hints for corotating planes in two of these seven cases. In both cases, however, only three dwarfs have distance estimates and are consistent with being satellites of the central ETG and have velocity measurements. In order to make robust claims about potential phase-space correlations in these structures, a higher number of velocity measurements will be necessary.

\section{Conclusions}
\label{sec:conclusions}

In the present work, we analyzed the 2D spatial distribution of a catalog consisting of 2210 dwarf galaxies around the ETGs targeted by the MATLAS survey in low- and moderate- density environments. We used two different automatic detection methods to identify any potential flattened structures in the 150 fields. Both the Hough technique and the TLS fitting method were used to derive the best fit to the dwarfs in each field. A comparison of the PAs resulting from the two methods revealed an agreement of $\sim$\,81\,\%. For subsequent analysis we removed fields with a significant overlap with others and only use fields in which the two methods agree. This cut leaves 119 fields (67 feature a single ETG). \par
Using TLS fitting, including recursive outlier removal and MC simulations, we identified 31 structures at the 2$\sigma$ confidence level. These automatically identified flattened structures show a distribution for the rms length with a maximum at around 150\,-\,180\,kpc. The rms height shows a pronounced maximum at $\sim$\,25\,kpc. The majority of the structures detected in this work show dimensions between those of the VPOS or GPoA and the CASP. The axis ratio (rms height over rms length) shows a flat distribution with an equal number of fields in the bins 0.1\,-\,0.25 and 0.35\,-\,0.4. The axis ratios of the structures proposed in this work agree well with the ones from the best studied planes in the Local Volume. Estimation of the virial radii of the massive host galaxies reveals that all of the flattened structures reside within this boundary. \par
Comparison of the photometric and kinematic position angles of the hosts with the orientation of the flattened structures reveals no preferred arrangement. Under the assumption that the orientation of the massive galaxies in the field hint at the large-scale structure, we find that 50\,\% of the flattened structure's normal vectors in fields with multiple massive galaxies avoid this orientation by $\geq$ 60$^{\circ}$, consistent with the results found in \citet{libeskind2019orientation}. More data on the orientations of cosmic filaments are needed to make conclusions about the alignment of these flattened structures. \par
We find no correlations when comparing the prevalence of significant flattened structures with the halo mass of isolated host ETGs or features related to merger history. There appears to be some excess of flattened structures around slow rotators (40\%) compared to fast rotators (16\%). A larger sample of galaxies featuring slow rotation, tails, streams, and shells is necessary to further test predictions. An analysis to examine corotation in these structures revealed a potential velocity trend in two out of seven testable cases based on the recessional velocities of dwarfs on either side of the massive host. In a follow-up paper, we will investigate the most confident flattened structures in detail and closely consider environmental features such as multiple massive hosts in the field, background galaxies, light contamination by stellar halos, and extended cirrus emission. \par
Since distance measurements for all dwarfs are not available at this time, our identification of flattened structures is limited to a 2D analysis. An analysis with kinematic data and cosmological simulations for a comparison with the $\Lambda$CDM model will shed more light on the satellite plane problem in more distant and low-to-medium density environments, which is a subject of future work. 

\begin{acknowledgements}
We thank the referee for the constructive report, which helped to clarify and improve the manuscript. Based on observations obtained with MegaPrime/MegaCam, a joint project of CFHT and CEA/IRFU, at the Canada-France-Hawaii Telescope (CFHT) which is operated by the National Research Council (NRC) of Canada, the Institut National des Science de l'Univers of the Centre National de la Recherche Scientifique (CNRS) of France, and the University of Hawaii. This work is based in part on data products produced at Terapix available at the Canadian Astronomy Data Centre as part of the Canada-France-Hawaii Telescope Legacy Survey, a collaborative project of NRC and CNRS. N.H. acknowledges the Vice Rector for Research of the University of Innsbruck for the granted scholarship. O.M. is grateful to the Swiss National Science Foundation for financial support. M.P. acknowledges the Vice Rector for Research of the University of Innsbruck for the granted scholarship. S.P. acknowledges support from the New Researcher Program (Shinjin grant No. 2019R1C1C1009600) through the National Research Foundation of Korea.
\end{acknowledgements}

\bibliographystyle{aa}
\bibliography{aanda}

\clearpage

\appendix
\section{Supplementary materials}

\subsection{Contaminated fields}
\label{appendix:1}

\begin{figure*}[htb!]
  \centering
    \includegraphics[width=0.49\linewidth]{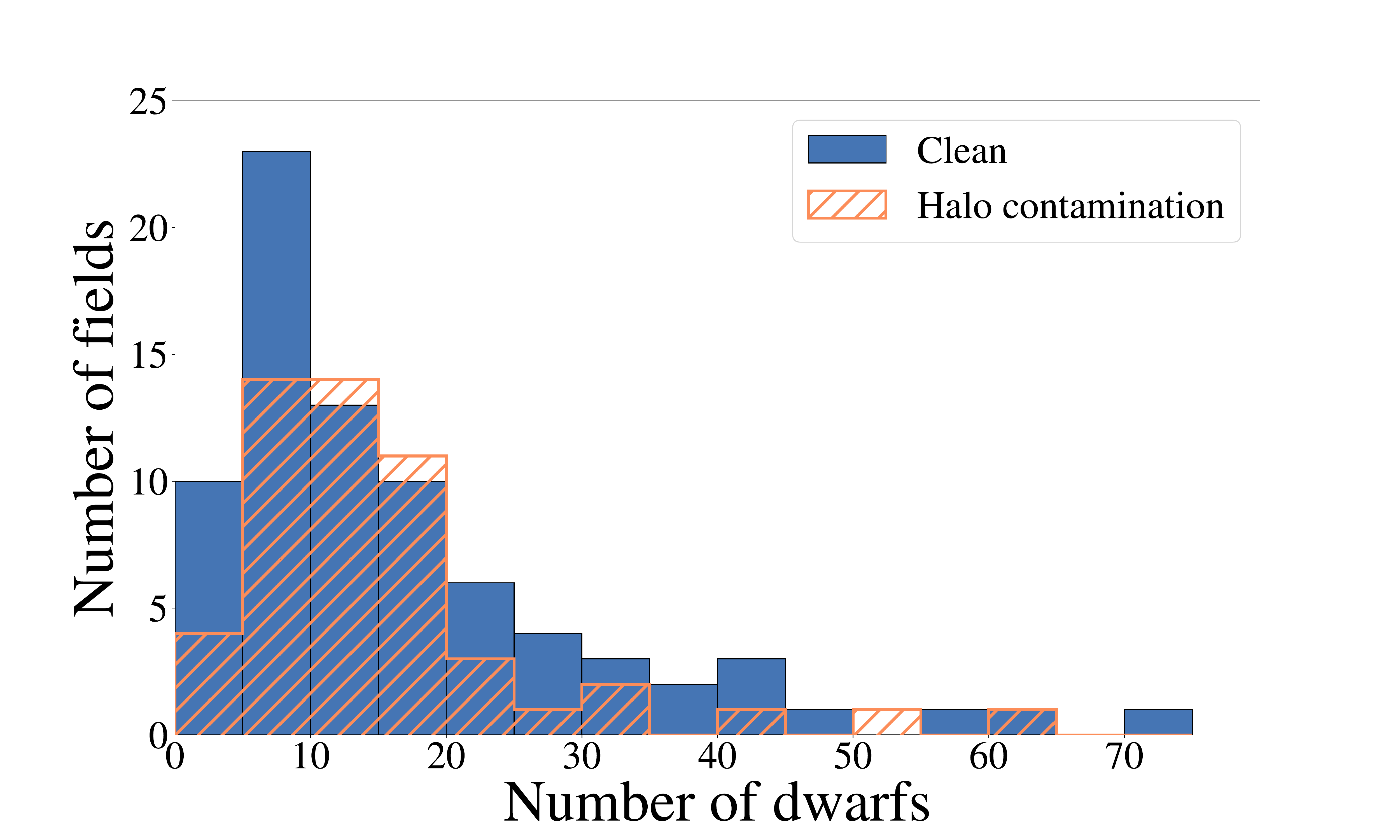}
    \includegraphics[width=0.49\linewidth]{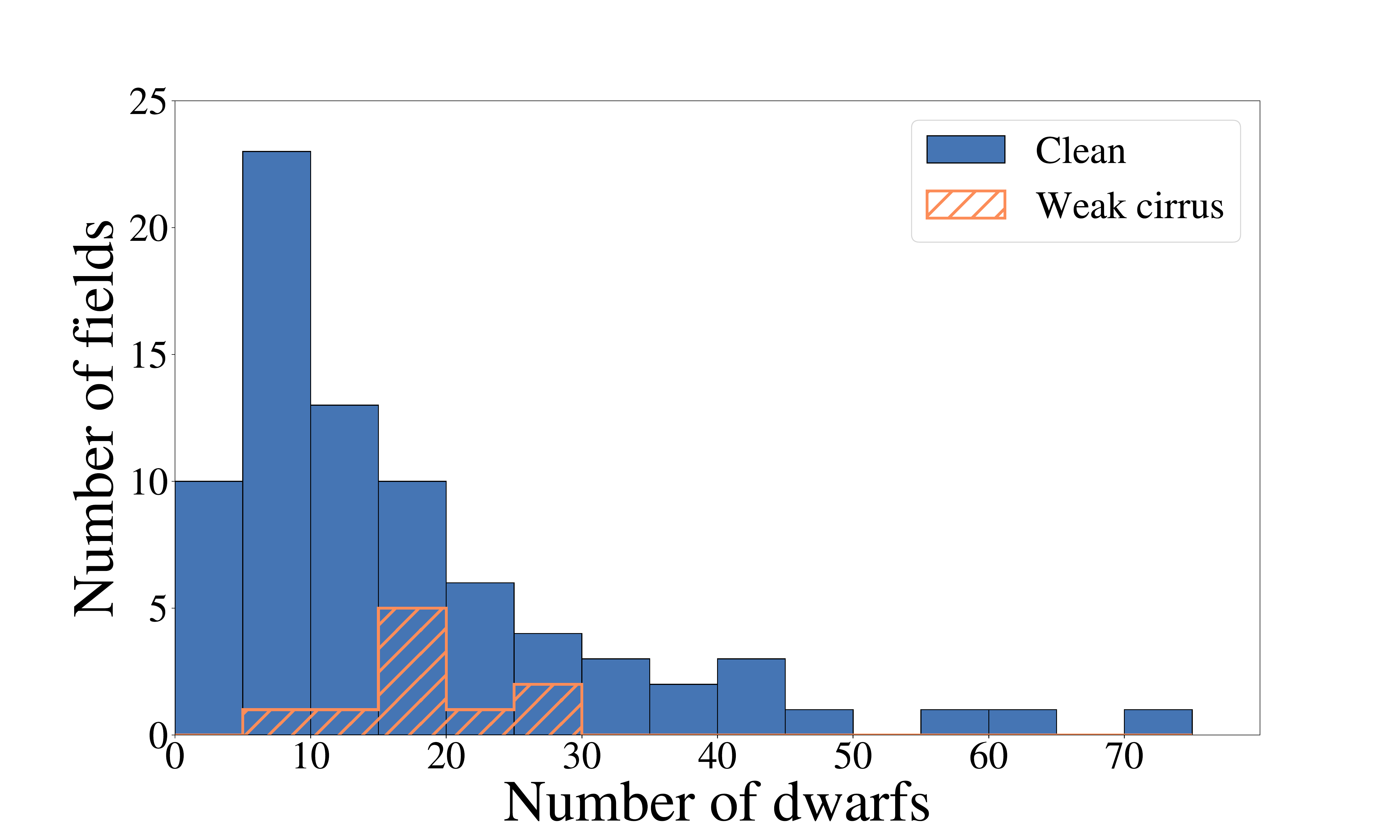}
    \includegraphics[width=0.49\linewidth]{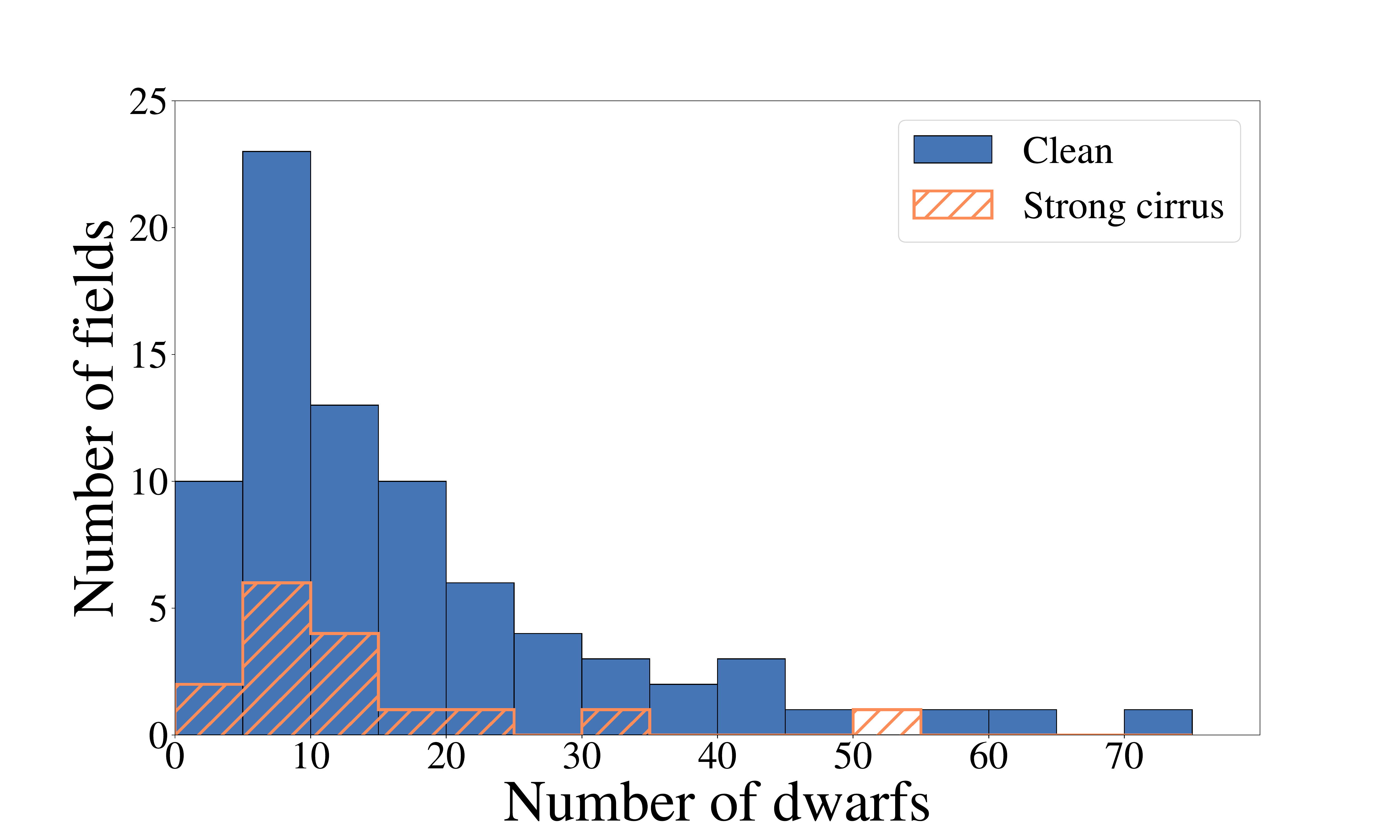}
    \includegraphics[width=0.49\linewidth]{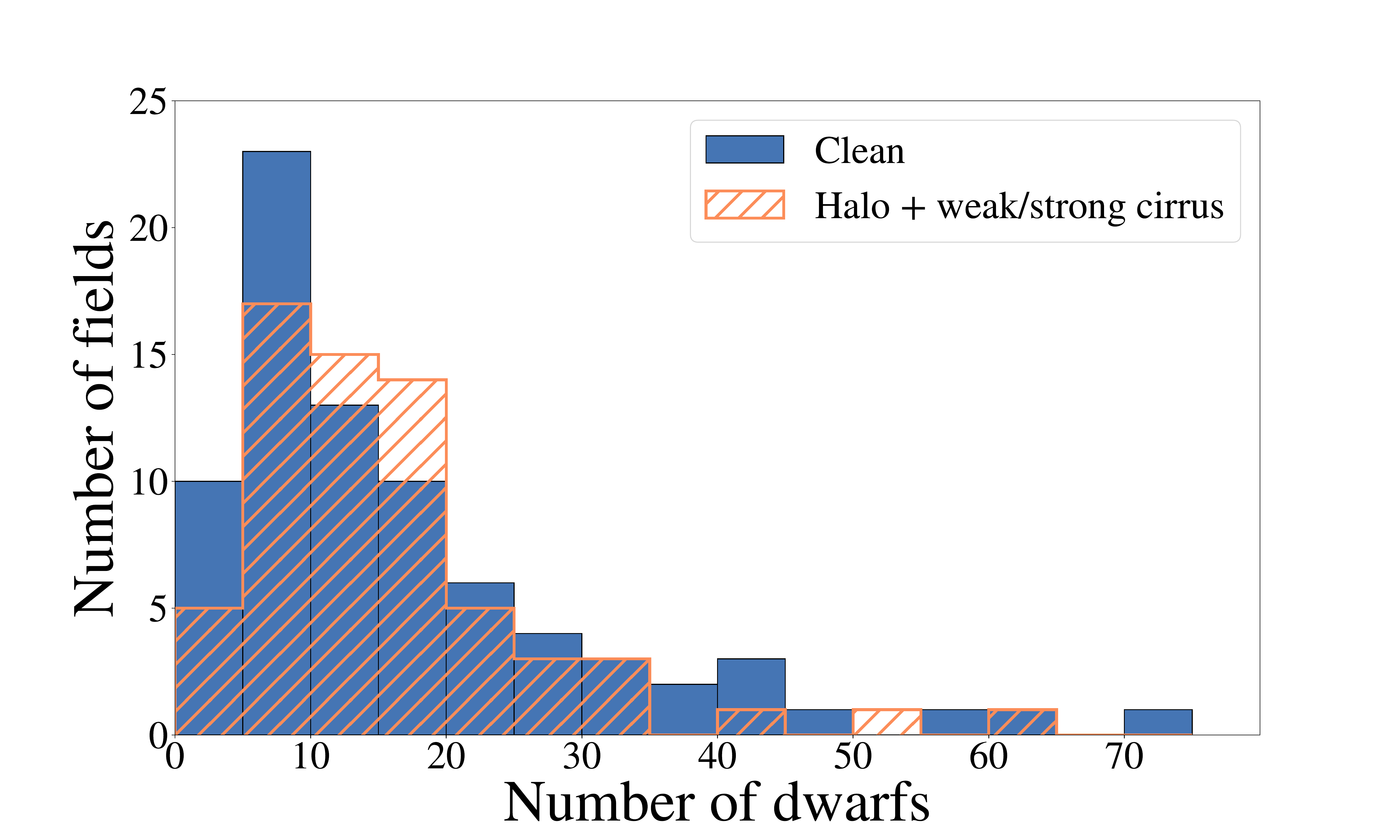}
  \caption{Number distributions of dwarfs detected in clean fields (blue bars) and in fields affected (orange dashed) by halos (top left), weak cirrus (top right), strong cirrus (bottom left), and all forms of contamination (bottom right).}
  \label{figure:cont_sep}
\end{figure*}

In order to estimate the effects of light contamination by bright stellar halos or extended cirrus emission, we plotted the number distributions of affected and unaffected fields. In \citet{bilek2020census} the authors split the contamination into three categories: halos, weak cirrus, and prominent or strong cirrus. We created four different plots where we examined the effects of halos, weak cirrus, strong cirrus, and all sources of contamination combined. The results are shown in Fig. \ref{figure:cont_sep}. A two-sample KS test comparing the number distributions of the clean and contaminated fields yields a p-value of $p_{C}$ = 0.79. We therefore cannot reject the null hypothesis that the two samples are drawn from the same parent distribution.

\subsection{Hough fitting technique}
\label{appendix:2}

The Hough fitting technique is designed to identify straight lines in images based on a voting process (described below) in which data points in the image space are assigned lines in a slope and intercept parameter space. In the simplest case, points are ordered in a straight line in a x-y coordinate system. Initially, ranges of possible intercepts and slopes are defined, of which one parameter pair would ideally describe a line through all points. For each of the points in the image space a range of possible intercepts and slopes is calculated and plotted in the parameter space. The location in parameter space where all lines cross reveals the pair of slope and intercept of the fit line that describes the linear relationship of the points in image space. \par 

Based on this idea \citet{duda1971use} further explored the method, realizing that the initial slope and intercept parameter space had no bounds, which poses a problem for computation. Instead they transformed the data points to a different parameter space. The line equation in this new space is

\begin{equation}
\rho = x\,\cos(\theta) + y\,sin(\theta)
\label{hough_formula}
\end{equation}

\noindent with x and y the coordinates of the data points in normal image space and $\theta$ the angle between the line showing the orthogonal distance to the image origin and the x axis. The parameter $\rho$ is the normal distance from the fit line to the origin. In this way the parameter space is closed with $\theta \in [-\pi/2,\pi/2]$ and $\rho \in [-d, d]$, $d$ being the image diagonal. Arbitrary precision in the results can be achieved by increasing the number of points in the $(\theta,\rho)$-range. Following the method described by Hough, for every point in the image space all possible values for $\theta$ and $\rho$ in the specified parameter ranges are calculated. To automate and record this procedure, a so called Hough accumulator -- a matrix structure with dimensions according to the chosen number of values in the two parameter ranges -- is created. Every line corresponds to one value of the $\rho$-parameter and the columns to the $\theta$ values. Starting for the first data point in the image space, all values in the $\theta$ range are inserted into Eq. (\ref{hough_formula}) yielding the $\rho$ values. These are compared to the values in the chosen $\rho$ parameter range and at the index pair of the closest match of $\rho$ and the corresponding $\theta$, the initial value (0) is raised by 1. After repeating for all data points, the accumulator is searched for the index pair with the highest value. This $(\rho,\theta)$ pair has the most votes in this system and a back-transformation to the slope and intercept space yields the best fit line through the data points in the image. \par

\begin{figure}[t]
\centering
\includegraphics[width=\linewidth]{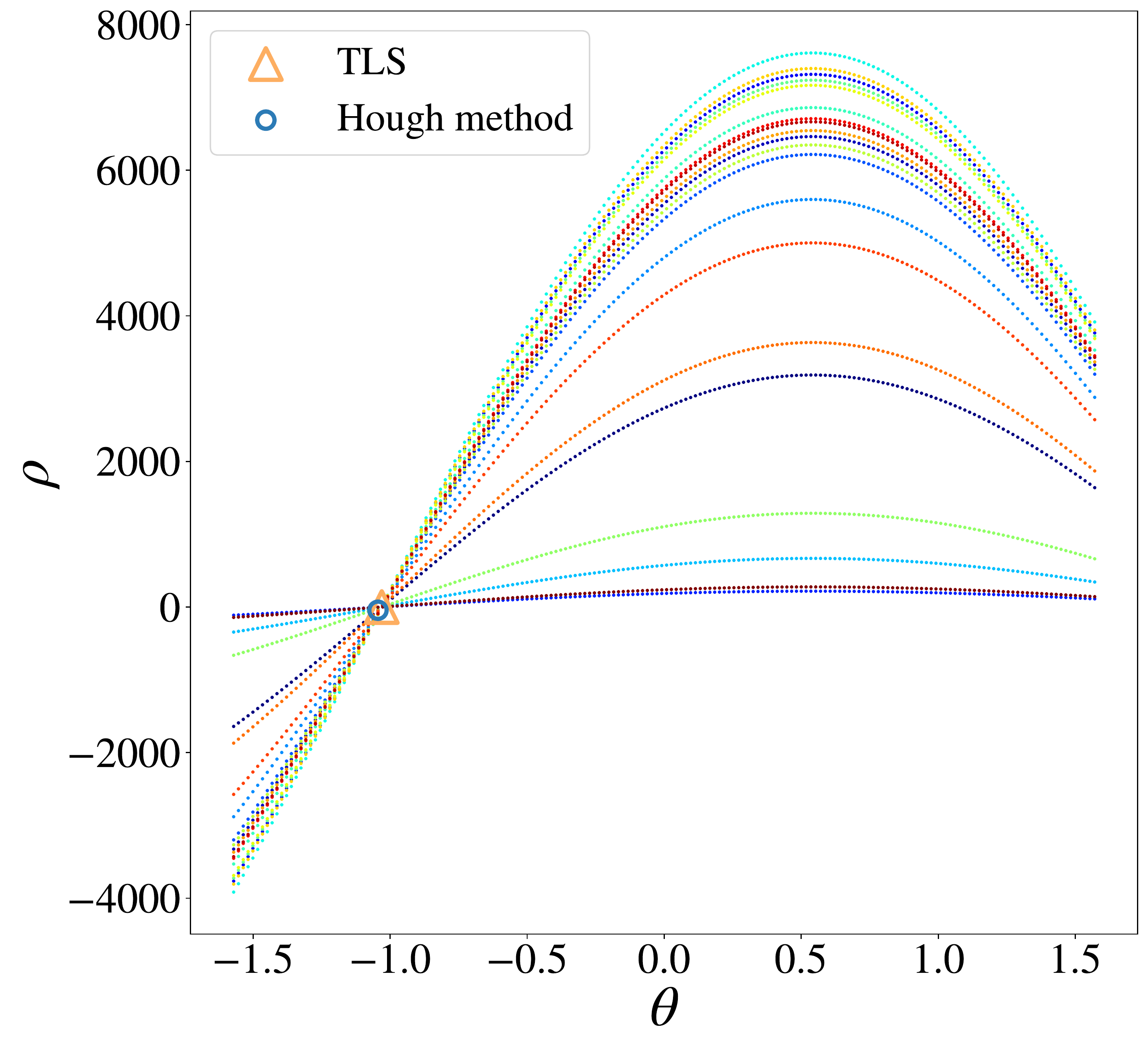}
\caption{Hough parameter space for the case of data points arranged in a perfect line. Each colored curve corresponds to one data point in the image space. There is exactly one point where all curves cross, marking the parameter pair which optimally describes the linear relationship. Both the Hough method (blue circle) and the TLS method (orange triangle) find this point.}
\label{figure:hough_perfect}
\end{figure}

In its original form, this method was developed for the detection of lines or fits through points arranged in a perfect line, thus it was not designed for this case where we are trying to fit a model to a scatter plot. Inspecting the accumulator searching for the highest value revealed that the votes were too spread out in the parameter space. Many $(\rho,\theta)$ pairs would get the same low number of votes and no clear preferred fit could be extracted this way. In order to visualize the details of the voting process, we plotted the votes of every data point in the parameter space. For every single data point in image space a curve (as opposed to a line in the case of slope and intercept parameter space) of points in parameter space is generated. The location where these curves cross, marks the pair of values that best fit the data points. For points arranged in a perfect line, there would be exactly one crossing with all curves involved. An example of this case is shown in Fig. \ref{figure:hough_perfect}. However, since our data points are scattered, the curves show many crossings, with the tighter the linear scatter the more concentrated the crossing region. Since choosing one crossing to describe the behavior of all points appeared too arbitrary after visual testing, we searched within a user-defined region around every point in the parameter space, in order to maximize the number of curves that fall within the region of interest. The goal was to find the parameter pair with the most points from different curves, in the defined search area around it. The size of this search region can be adjusted in order to search for different degrees of scatter. This method will find the most clustered substructure on the user-defined scales. \par

Since visually there appear to be flattened structures in a range of different scales in our data, we adopted a method to assess the optimal size of the Hough search region in each field individually. We scanned each field with ten equidistant search area sizes, selected such that the smallest boxes would identify the GPoA and the thinnest visual structures in our data, the CASP would be picked out with a box near the center of the range, and the thickest of our proposed flattened structures that contain a large number of dwarfs are detected by the largest box. The optimal box selection is based on the assumption that the ideal flattened structure would be one where flattening ($c/a$) and number of dwarf members are maximized simultaneously. We therefore scanned over all ten box sizes and compared the relationship between Hough voters and the flatness of the resulting structures. Testing revealed that simply dividing the number of Hough voters by the flatness results in the flatness dominating the ratio, such that the smallest box is picked out in every case. To combat this, the number of Hough voters was squared and then divided by the flatness. This way the numerator gains sufficient weight such that a range of different boxes are picked out across the 150 fields in our data. The size of the search area resulting in the maximum of these ratios was picked out in each field.

\begin{figure}[htbp!]
\centering
\begin{subfigure}[t]{\linewidth}
   \includegraphics[width=\linewidth]{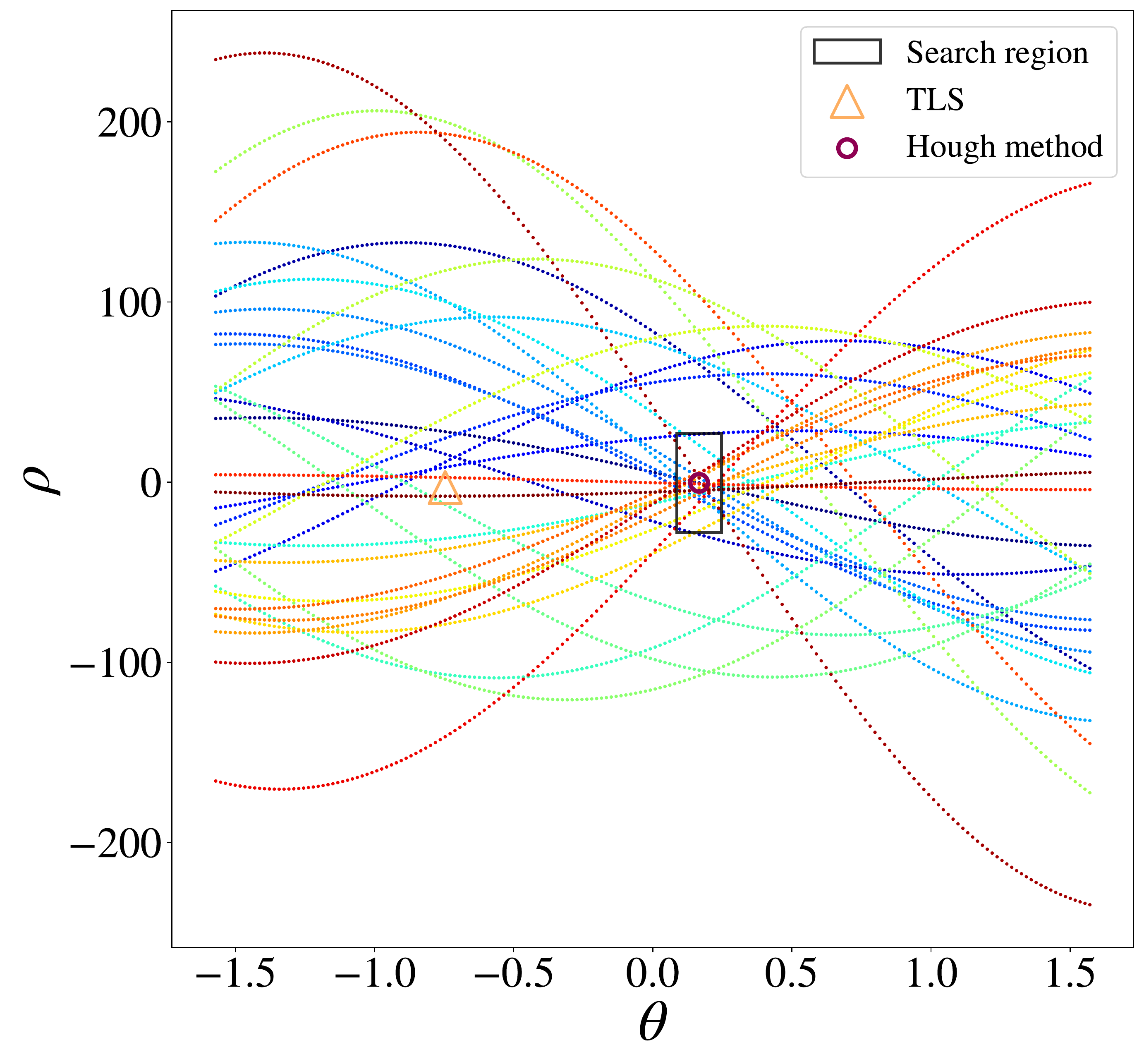}
   \caption{}
   \label{figure:hough_space_M31} 
\end{subfigure}
\begin{subfigure}[t]{\linewidth}
   \includegraphics[width=\linewidth]{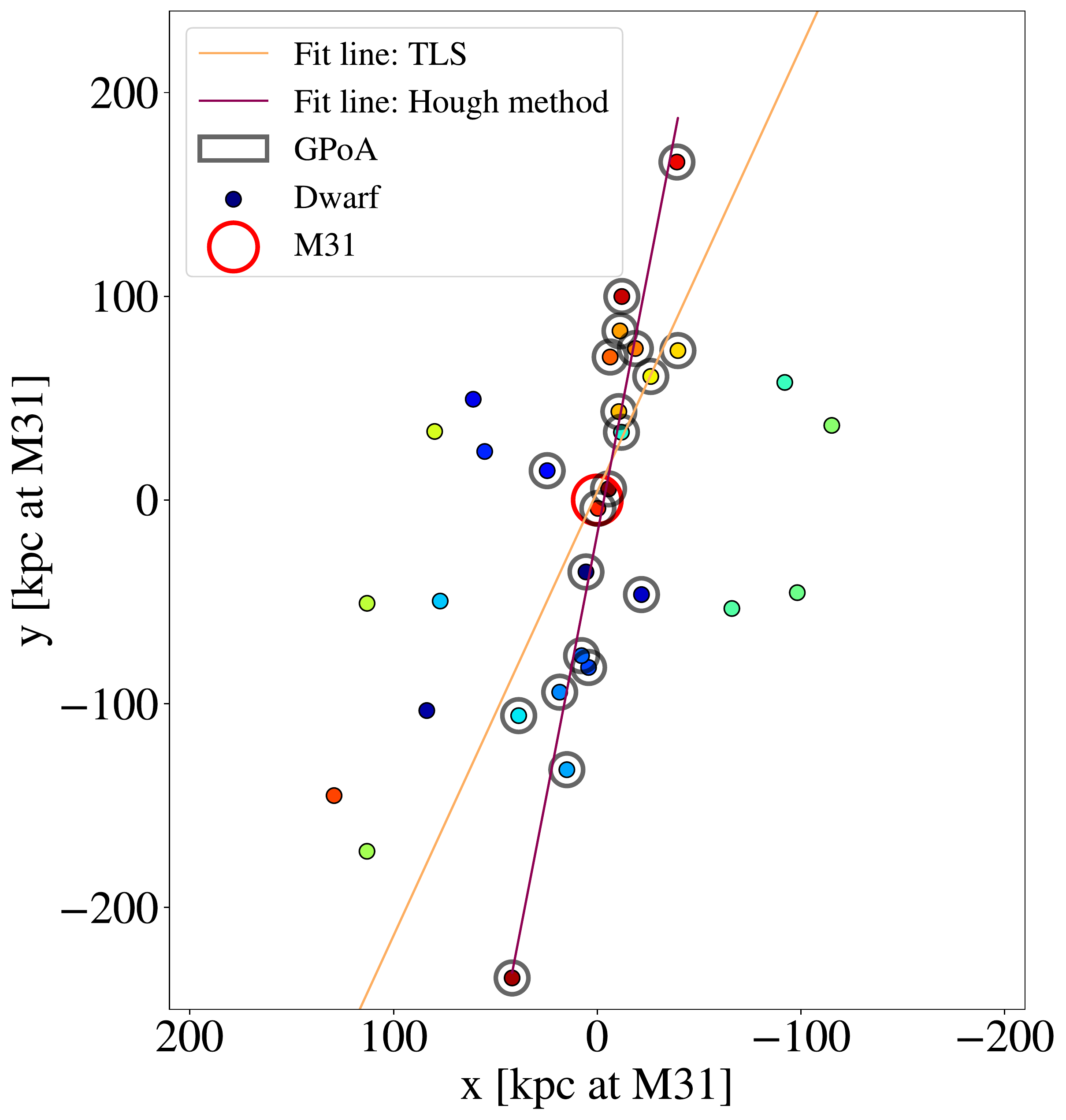}
   \caption{}
   \label{figure:hough_scatter_M31}
\end{subfigure}

\caption[]{(a) M31 dwarfs in the Hough parameter space. Each curve results from the calculation of parameters using Eq. (\ref{hough_formula}). The densest crossing region shows the parameter pair fitting the majority of the data. The black rectangle shows the size of the search area. The magenta circle points to the parameter pair which best describes the clustered substructure. The orange triangle shows the parameter location of a TLS fit on the data points. (b) Scatter plot showing the M31 (red circle) satellite system in right ascension and declination. The orange line shows a TLS fit on the data points. The magenta line is the fit produced by the Hough method. The dwarfs circled in gray are members of the GPoA and at the same time voted for the best fit parameters. The colors of the data points correspond to the ones of the lines in plot (a).}
\label{figure:hough_M31}
\end{figure}

We tested this method on the dwarf satellite system around Andromeda, using data from \citet{ibata2013vast} and \citet{pawlowski2013dwarf}. We visualized the voting system in the Hough parameter space in Fig. \ref{figure:hough_space_M31}. Each colored curve of points is the result of the voting process of a single data point. In this particular example, the densest crossing region is pronounced and isolated from the rest of the voting curves, so the location of the best fit parameters can be guessed visually. The search area marked as the black box was picked out from a range of ten user-defined sizes following the above described process. In the case of the GPoA the smallest possible search area is used. The magenta circle points to the parameter pair at the center of the black box, while the orange triangle shows the parameters according to a TLS fit on the data points. In Fig. \ref{figure:hough_scatter_M31} we show a scatter plot of the M31 satellite system. The colored dots show the data sample used in \citet{ibata2013vast} and five additional dwarfs used in \citet{pawlowski2013dwarf}. The red circle near the center shows the location of M31. The magenta fit line corresponds to the parameter pair marked as the magenta circle in Fig. \ref{figure:hough_M31} and the dwarfs circled in gray passed the densest region inside the black box in the voting process. These dwarf galaxies are members of the GPoA. As a comparison we added the orange TLS fit line to this data set. \par

\begin{figure}[htbp!]
\centering
\begin{subfigure}[t]{\linewidth}
   \includegraphics[width=\linewidth]{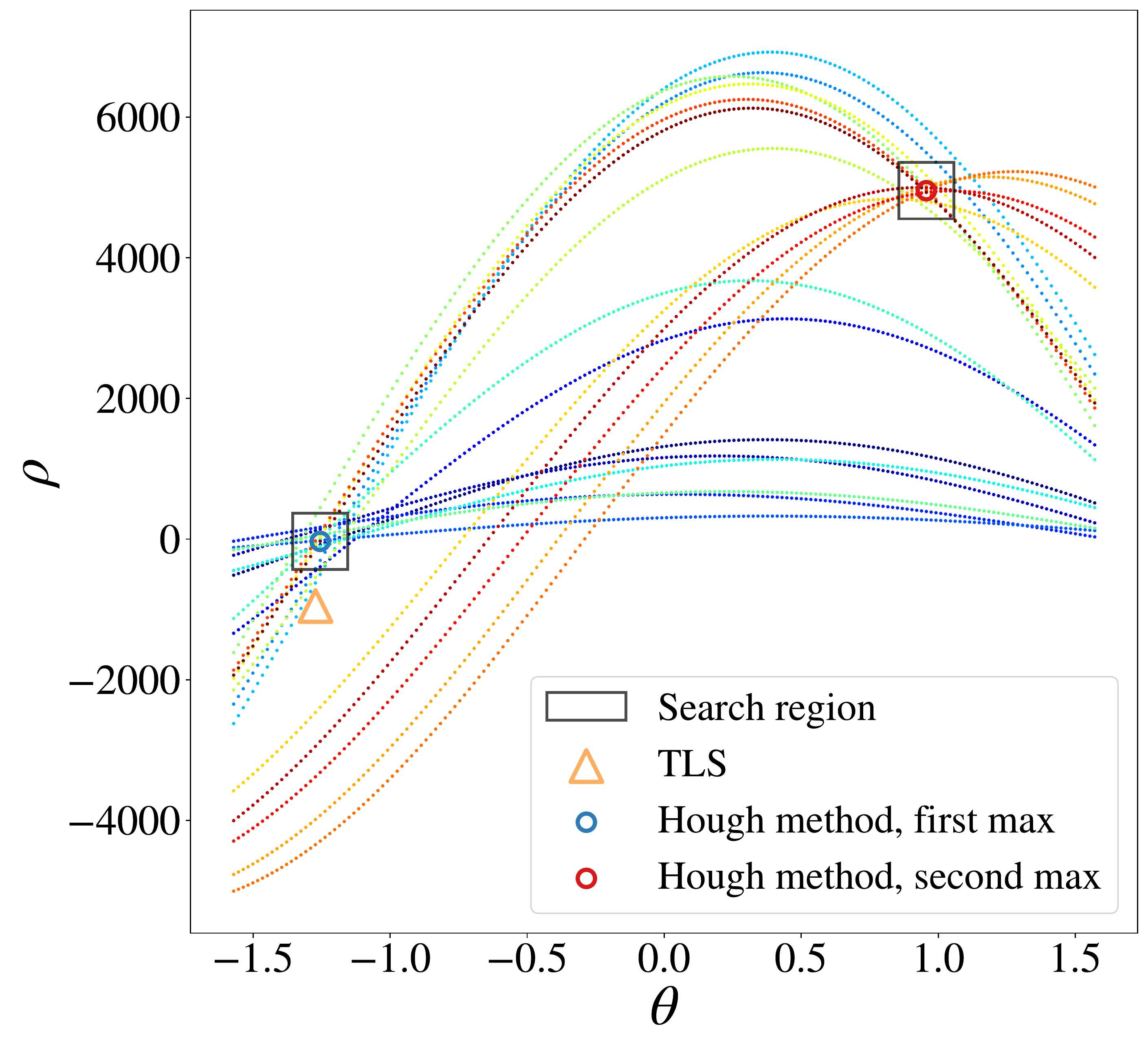}
   \caption{}
   \label{figure:hough1} 
\end{subfigure}
\begin{subfigure}[t]{\linewidth}
   \includegraphics[width=\linewidth]{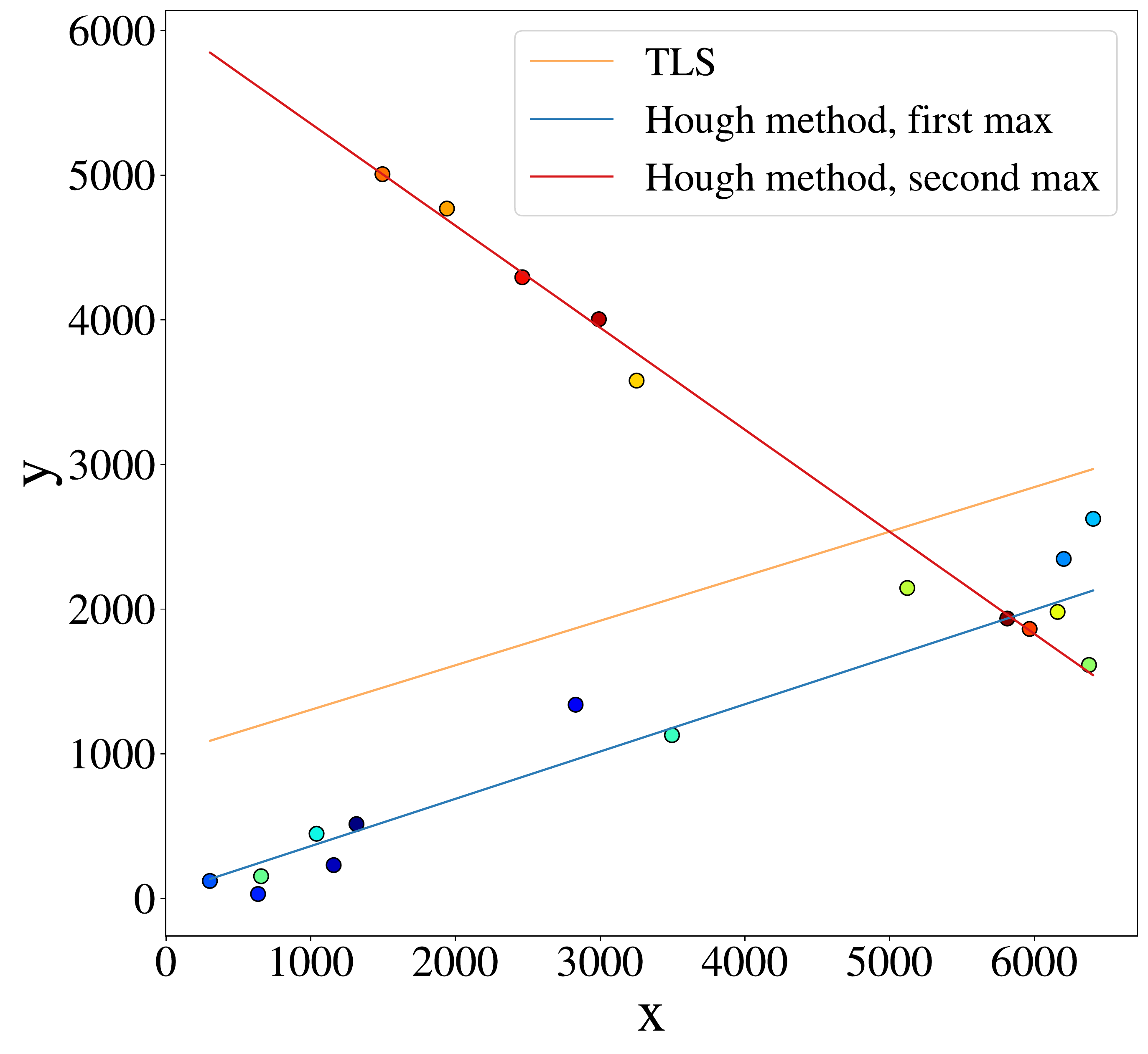}
   \caption{}
   \label{figure:hough2}
\end{subfigure}

\caption[]{(a) Curves corresponding to votes of data points in Hough space. Each curve results from the calculation of parameters using Eq. (\ref{hough_formula}). The areas with the most crossings show the parameter pairs fitting two linear scatters. The black rectangle shows the search area for the maxima. For every point the number of different neighbors within are counted. Each point can vote for the two different fits. The blue and red circle show the first and second maxima, respectively. The orange triangle shows the parameter location of a TLS fit on the data points. (b) Scatter plot showing two randomly generated linear scatters fitted via the Hough method. The blue and red lines show the first and second detected regions of maximal crossings in parameter space. The orange line shows a TLS fit on the data points. The colors of the data points correspond to the ones of the lines in plot (a).}
\label{figure:hough_example}
\end{figure}

We further added the ability to search for multiple maxima or multiple linear scatters in an image to the code. To ensure that different structures are detected, the points to be considered as a next maximum must be outside the search area of a prior maximum. Since in some cases two or more overlapping linear structures can be proposed from a visual inspection, this addition allowed us to compare the priority of the visual and automatic detections. In this work we only searched for a single maximum, that is a single flattened structure per field. This extension to multiple maxima could in principle be used in complicated cases with evidence for multiple flattened structures. Figure \ref{figure:hough1} illustrates the detection process after transformation into Hough space. In this case, 20 mock data points were arranged in two different linear scatters in a field. Each curve in Hough space again corresponds to one of the data points. The two locations where the most crossings occur are clearly visible. The black boxes show the search area in which the number of ($\theta, \rho$) belonging to different data points are counted. The blue and red circles show the first and second maximum found with this method while the orange triangle shows the parameter pair corresponding to a TLS fit, which considers all points in the image equally. Figure \ref{figure:hough2} shows the data points in x-y space and the fit lines corresponding to the first (blue) and second (red) maximum detected with the Hough method and the fit found with the TLS method (orange).

\subsection{Removed outliers}
\label{appendix:3}

\begin{figure*}[htb!]
  \centering
    \includegraphics[width=0.49\linewidth]{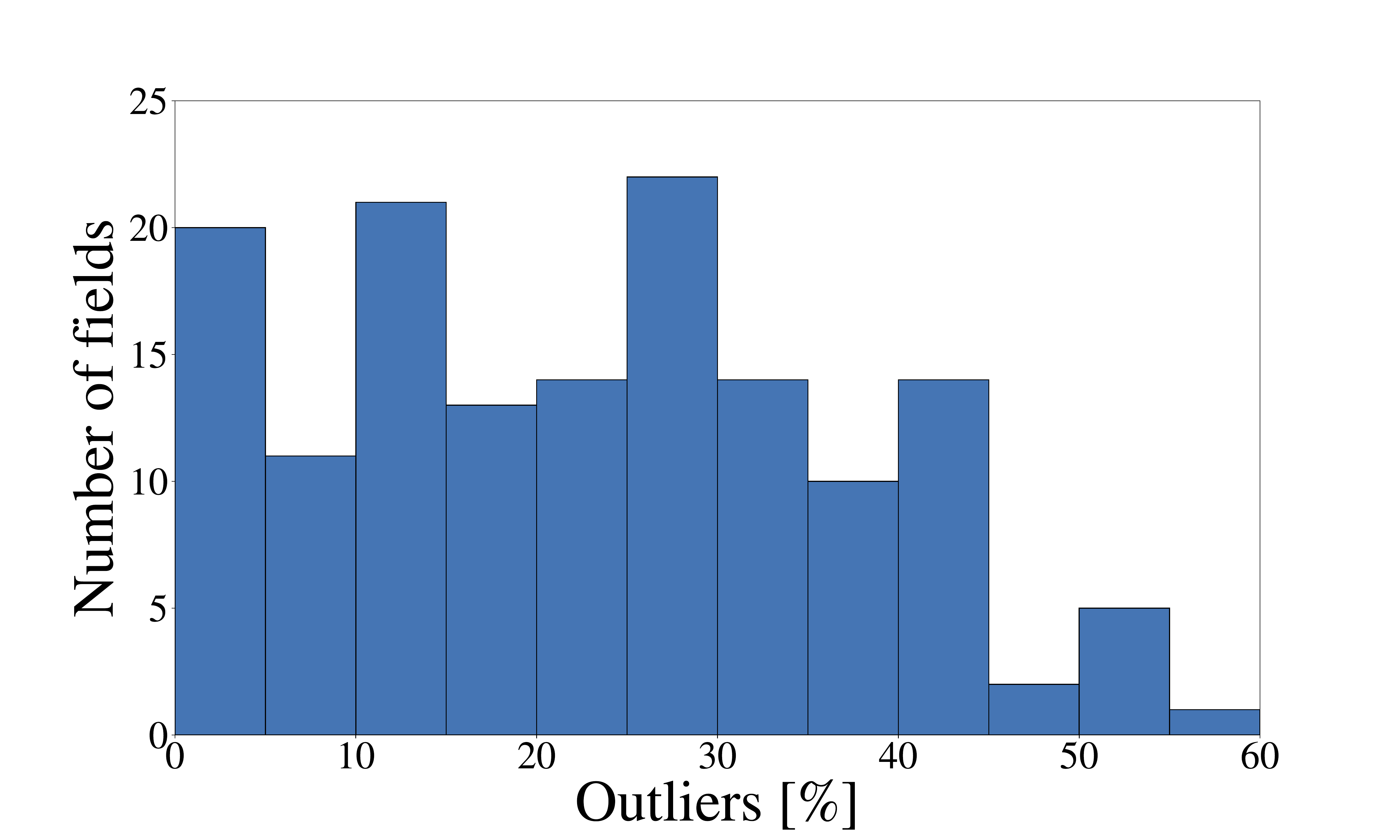}
    \includegraphics[width=0.49\linewidth]{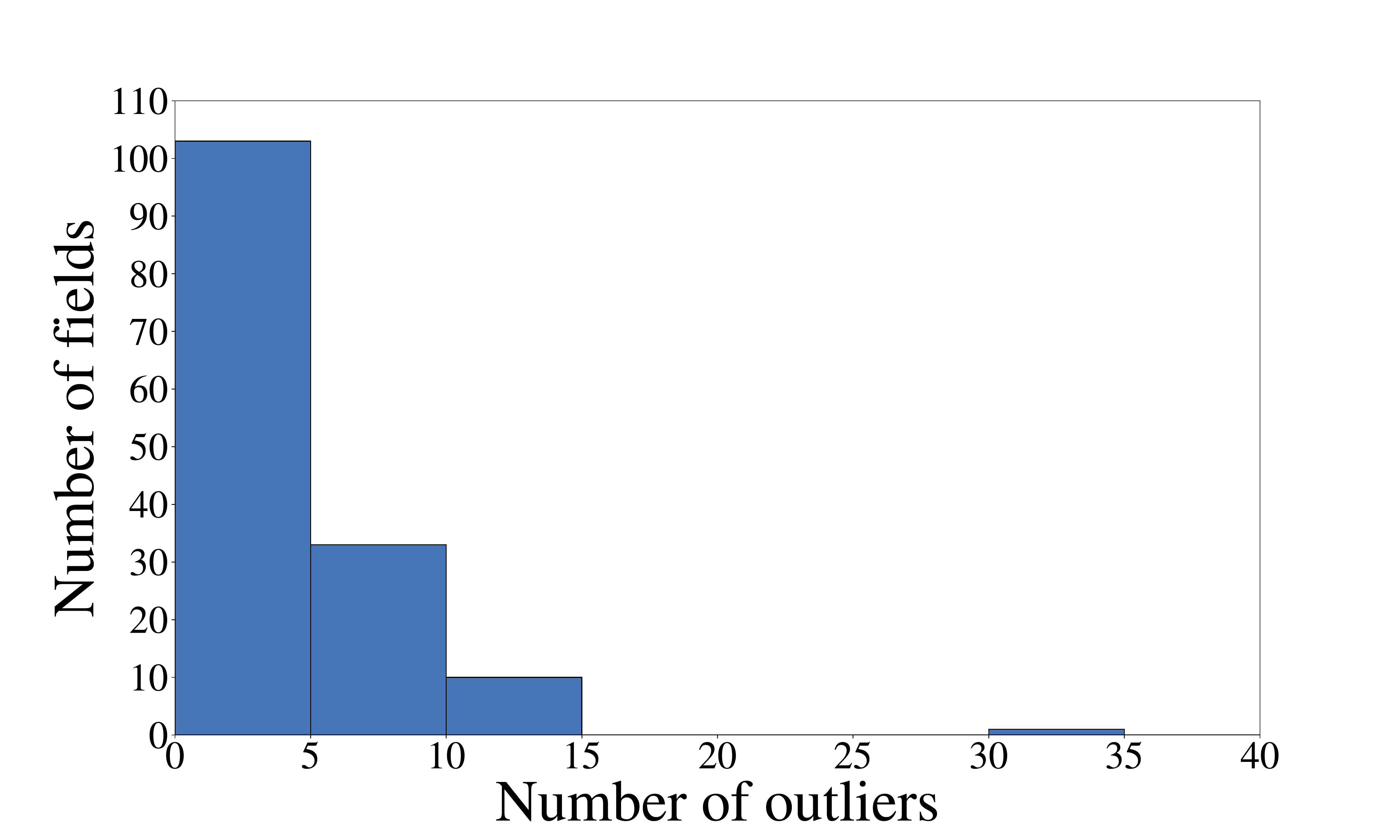}
    \includegraphics[width=0.49\linewidth]{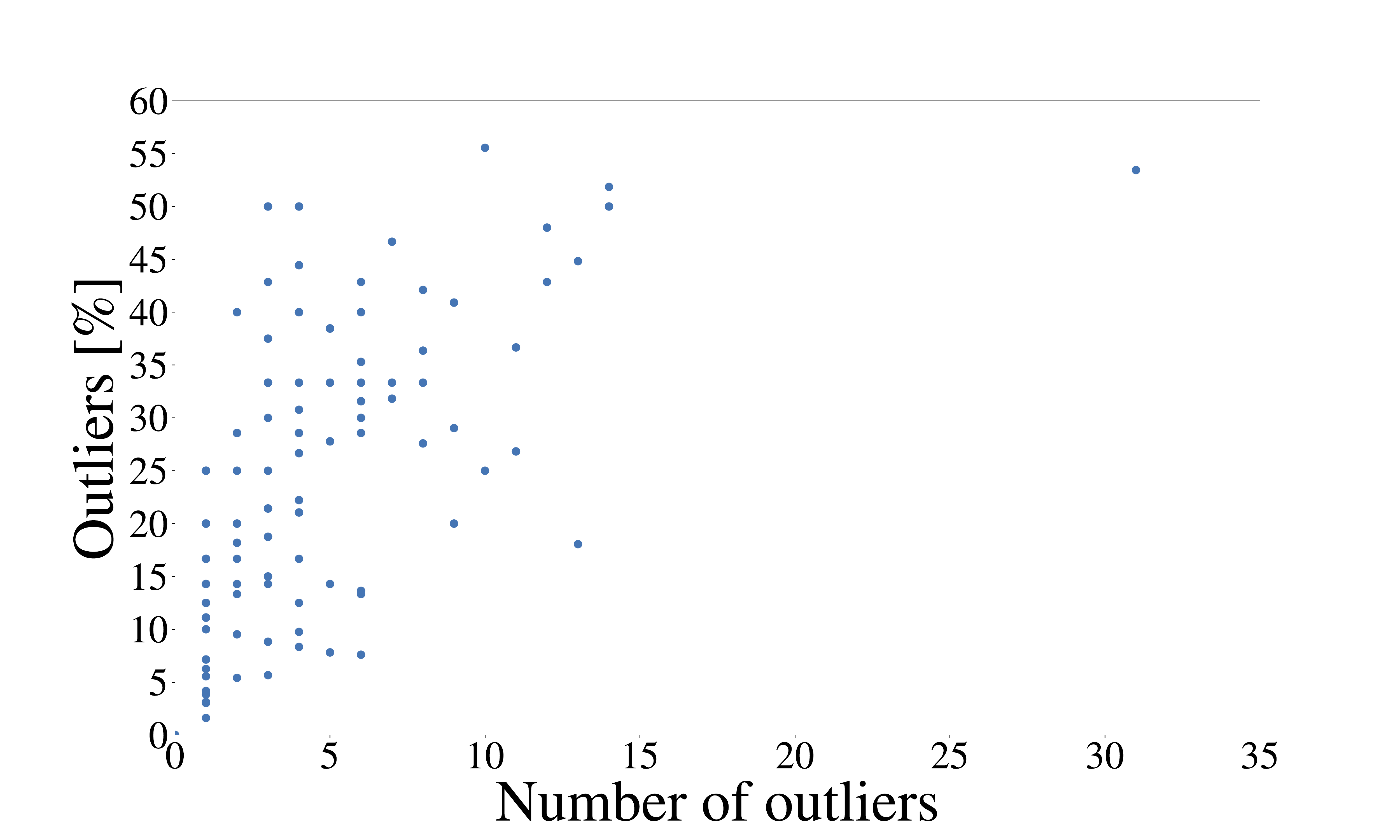}
  \caption{Number distributions of the removed outliers in percentile (top left) and absolute values (top right). The bottom center figure shows a comparison of the two.}
  \label{figure:outliers}
\end{figure*}

In this section we examine the distributions of removed outliers via the Hough method. The results for percentile and absolute numbers as well as a direct comparison between the two are shown in Fig. \ref{figure:outliers}.

\subsection{Field of view considerations}
\label{appendix:4}

\begin{figure}
\centering
\includegraphics[width=\linewidth]{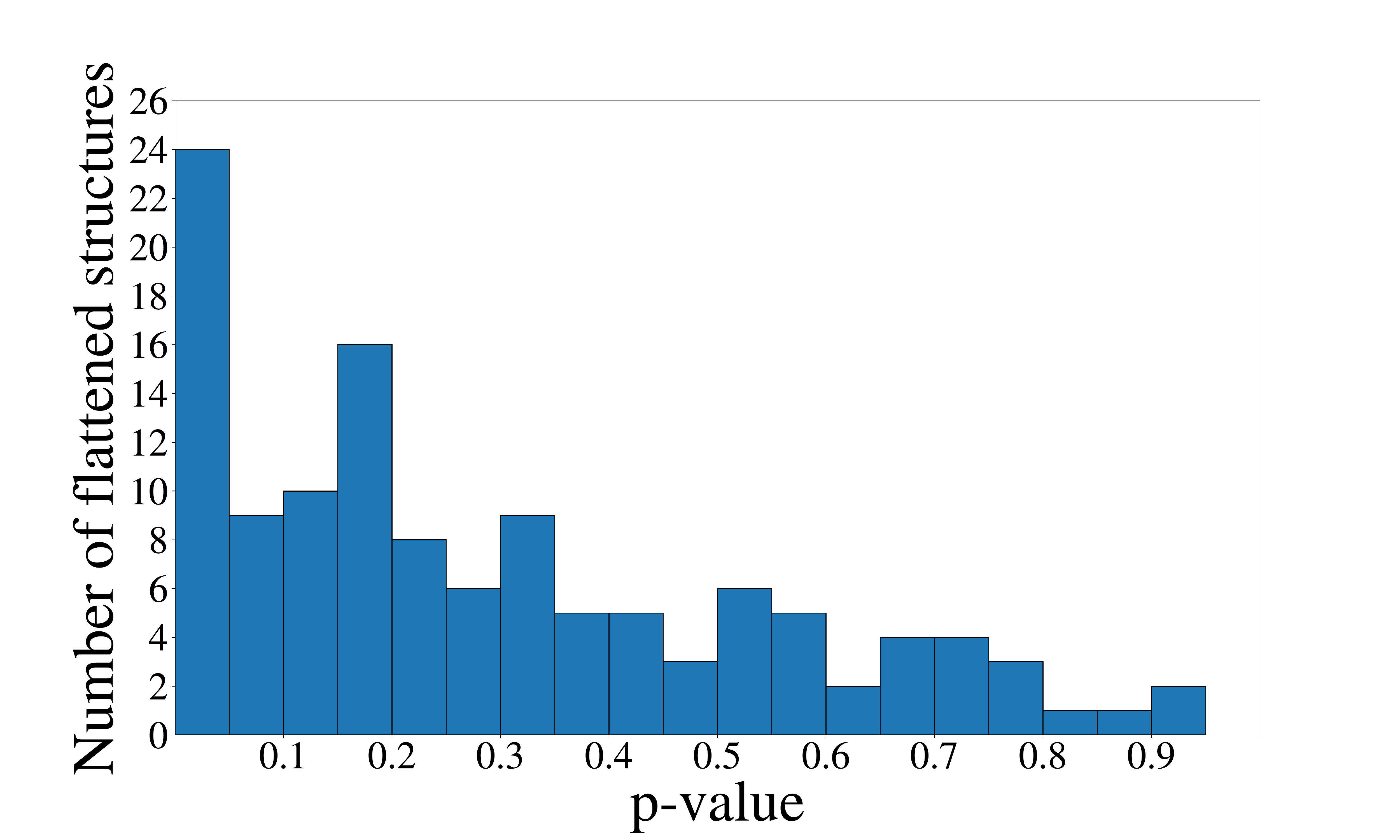}
\caption{P-value distribution of the dwarf structures in 123 fields. Only dwarfs in circular areas around the host galaxy were considered. Only unique fields (significantly overlapping fields excluded) where the two detection methods agree were included in this analysis.}
\label{figure:pvalFOV}
\end{figure}

\begin{figure}[htb]
\centering
\begin{subfigure}[t]{\linewidth}
   \includegraphics[width=\linewidth]{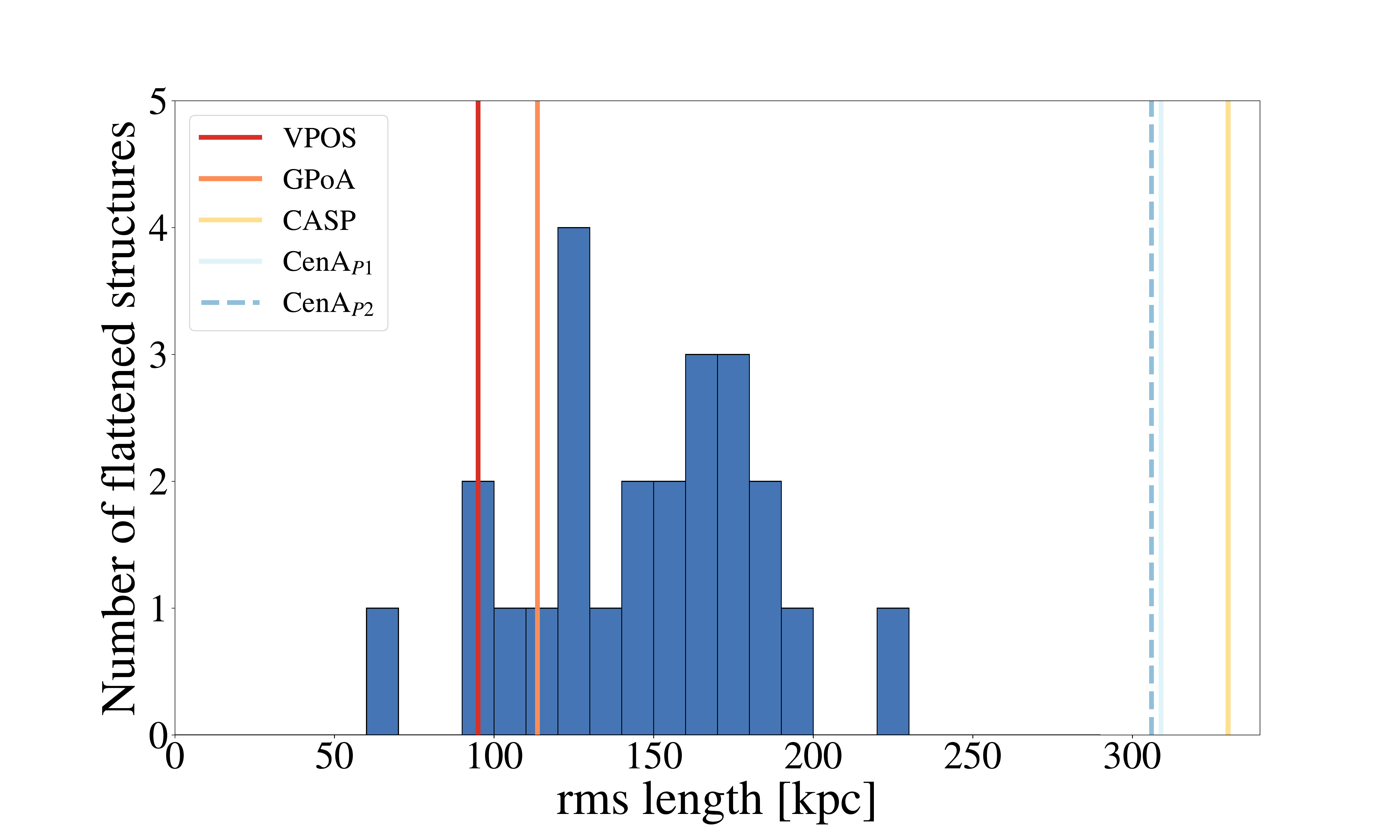}
   \caption{}
   \label{figure:dim1FOV}
\end{subfigure}
\begin{subfigure}[t]{\linewidth}
   \includegraphics[width=\linewidth]{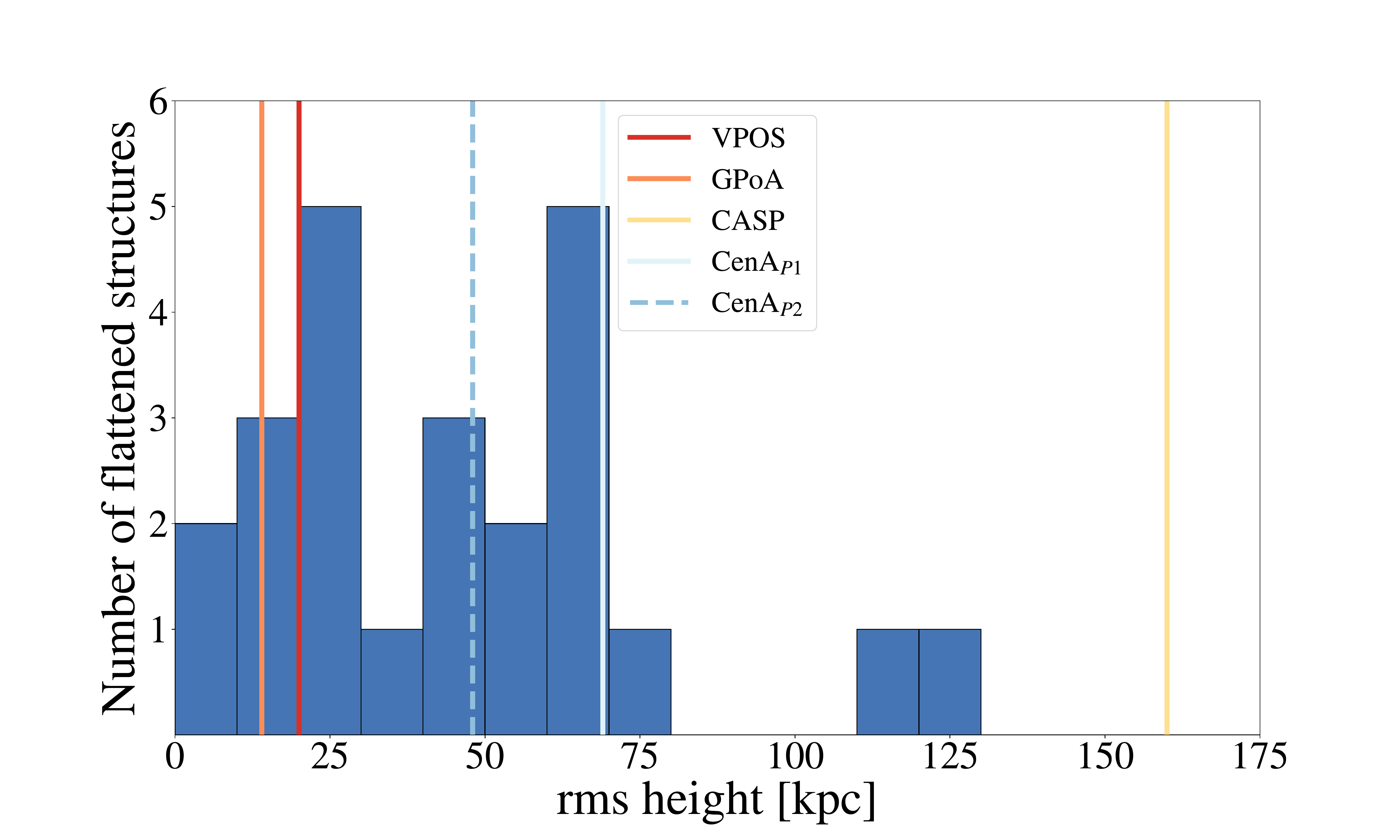}
   \caption{}
   \label{figure:dim2FOV}
\end{subfigure}
\begin{subfigure}[t]{\linewidth}
   \includegraphics[width=\linewidth]{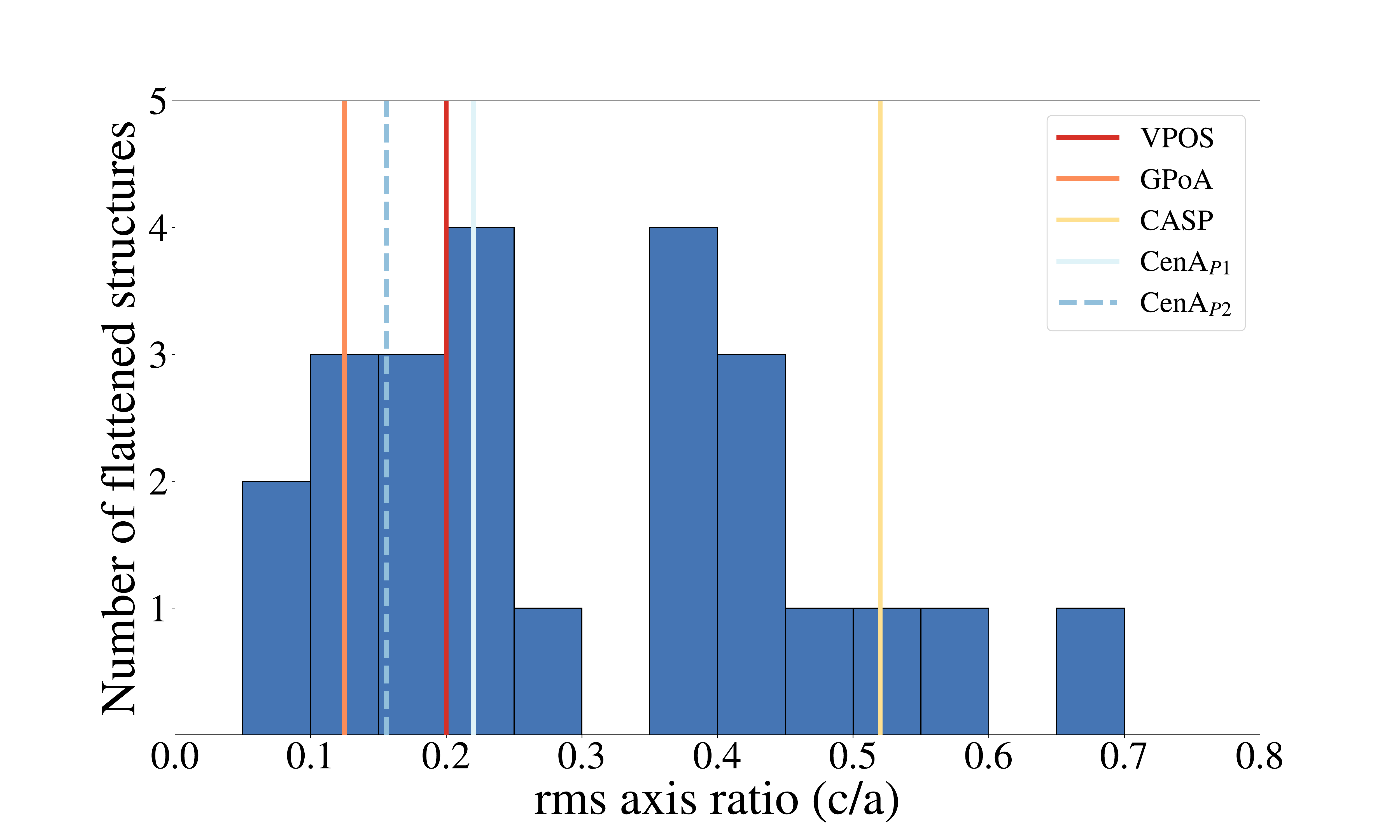}
   \caption{}
   \label{figure:dim3FOV}
\end{subfigure}

\caption[]{Rms dimensions of the automatically detected (p $\leq$ 0.05) structures in circular areas around the central host galaxy. These dimensions were calculated via the small angle approximation assuming all dwarfs are at the distance of the targeted host galaxy. Shown are the structure length (a), height (b), and axis ratio (c). Measured values of several well-studied planes are over plotted for comparison purposes. References for these are: \citet{pawlowski2013dwarf} (VPOS, GPoA), \citet{muller2018whirling} (CASP), \citet{tully2015two,muller2016testing} ($CenA_{P1}$, $CenA_{P2}$).}
\label{figure:dimensions_FOV}
\end{figure}

Our rectangular FOV makes diagonal best-fit lines more likely than other configurations. This bias can be addressed by only considering dwarfs within circular areas around the central ETG which we assumed to be the host galaxy. This restriction, however, leads to the exclusion of a considerable fraction of dwarfs ($\sim$\,27\%) near the corners of our fields which are instrumental for high statistical significance in many cases. Consequently, we only find 24 statistically significant flattened structures after removing these dwarfs on the outskirts of our fields. The resulting p-value distribution can be seen in Fig. \ref{figure:pvalFOV}. The rms dimensions are shown in Fig. \ref{figure:dimensions_FOV}.

\subsection{Structure dimensions with full dwarf sample}
\label{appendix:5}
\begin{figure}[htb]
\centering
\begin{subfigure}[t]{\linewidth}
   \includegraphics[width=\linewidth]{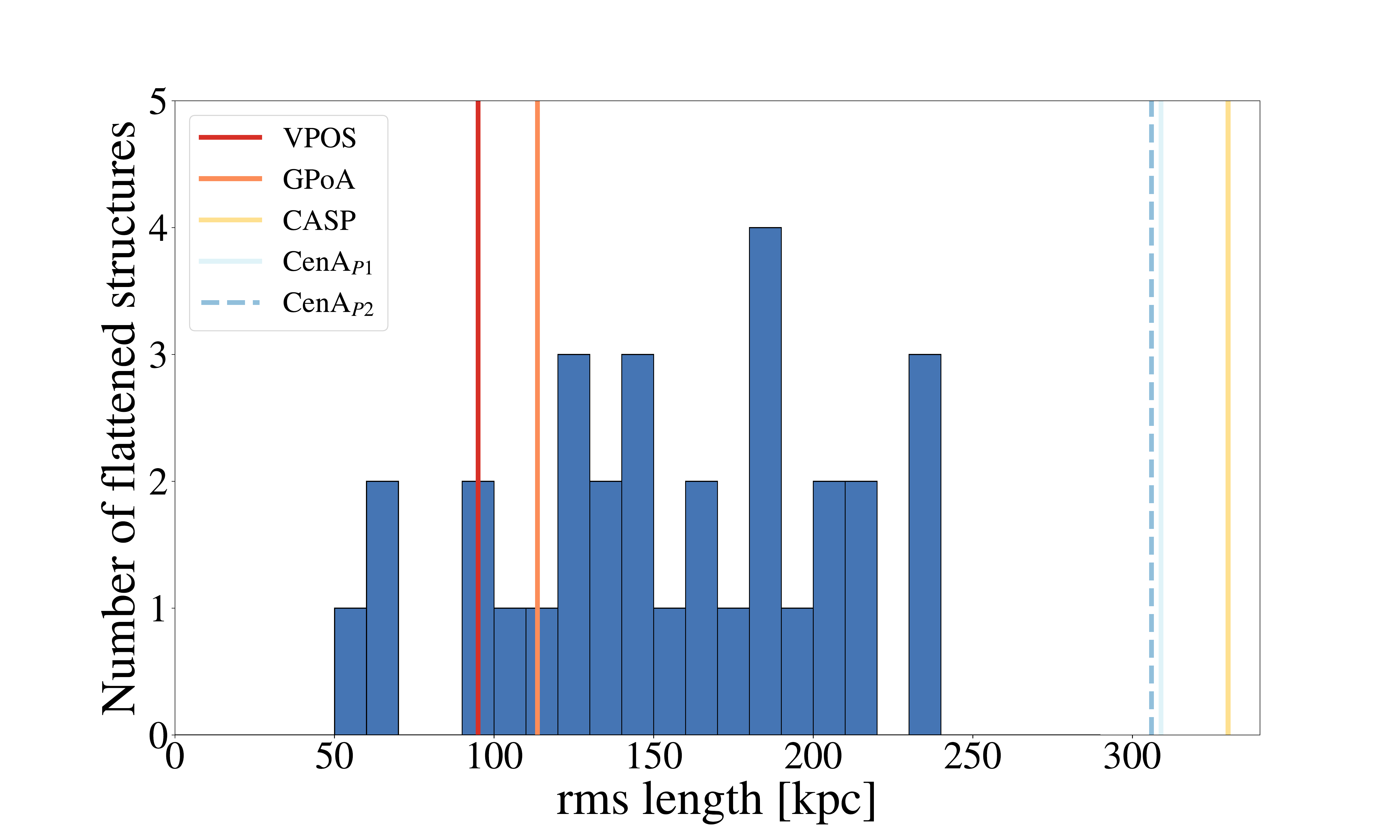}
   \caption{}
   \label{figure:dim1all} 
\end{subfigure}
\begin{subfigure}[t]{\linewidth}
   \includegraphics[width=\linewidth]{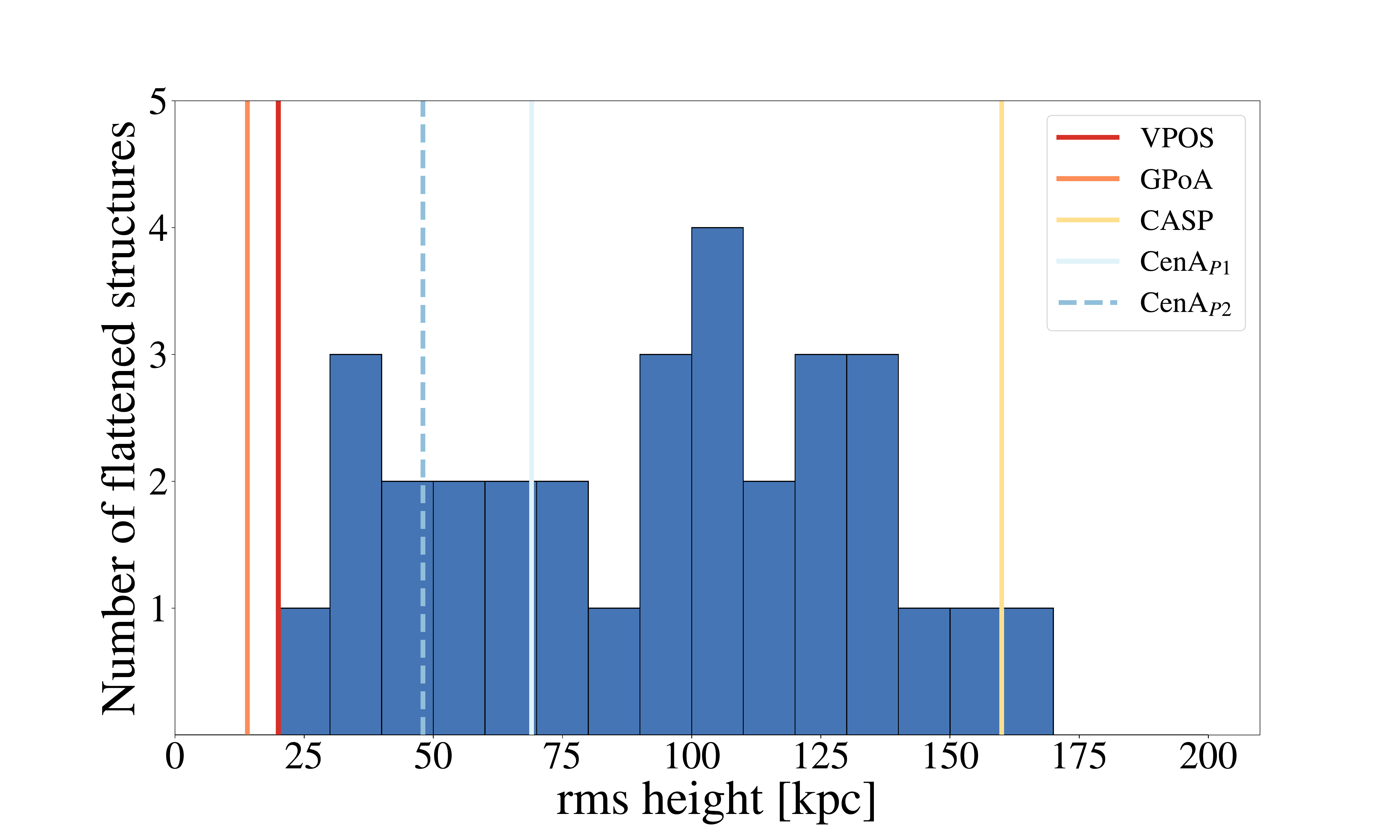}
   \caption{}
   \label{figure:dim2all}
\end{subfigure}
\begin{subfigure}[t]{\linewidth}
   \includegraphics[width=\linewidth]{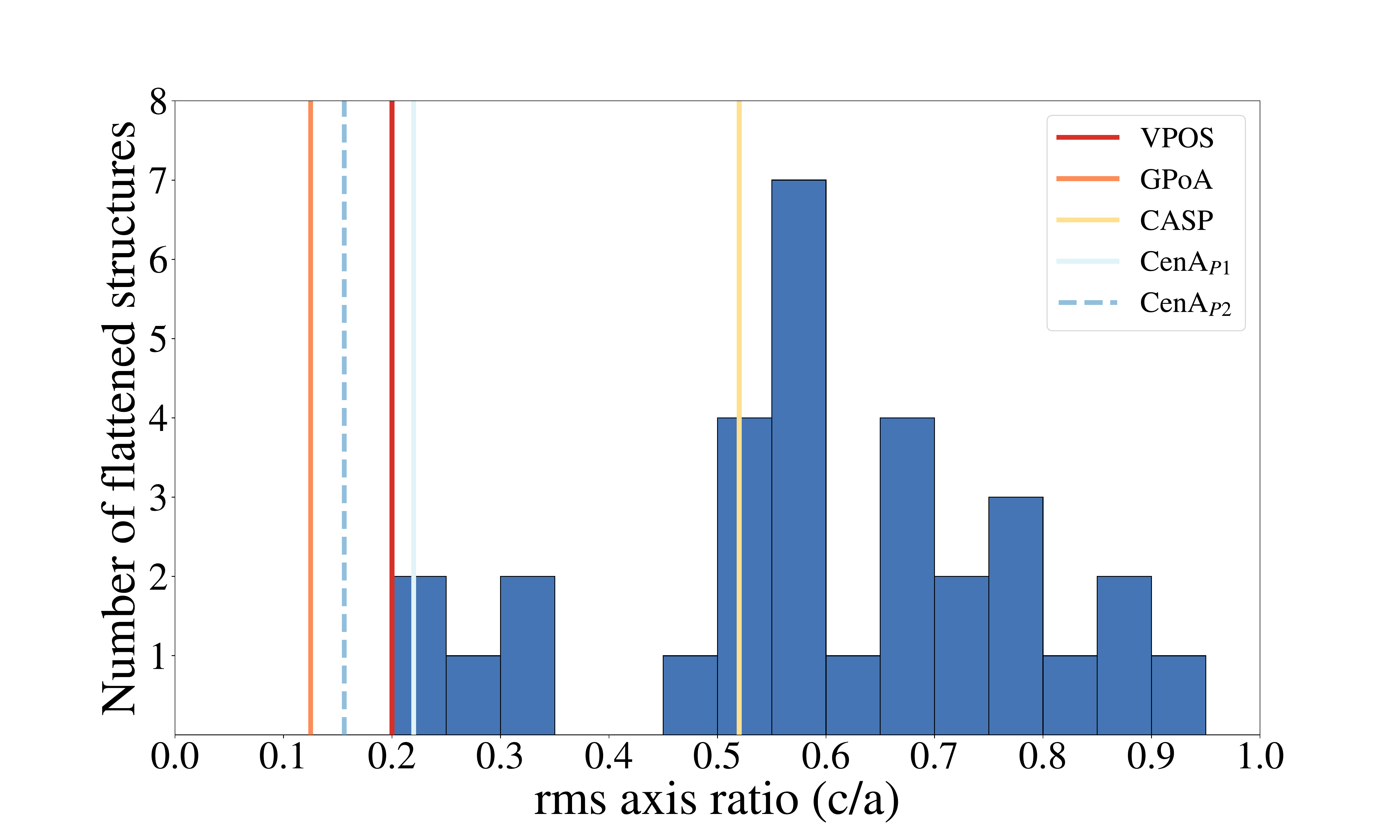}
   \caption{}
   \label{figure:dim3all}
\end{subfigure}

\caption[]{Rms dimensions of the automatically detected (p $\leq$ 0.05) structures using all dwarfs in the field. These dimensions were calculated via the small angle approximation assuming all dwarfs are at the distance of the targeted host galaxy. Shown are the structure length (a), height (b), and axis ratio (c). Measured values of several well-studied planes are over plotted for comparison purposes. References for these are: \citet{pawlowski2013dwarf} (VPOS, GPoA), \citet{muller2018whirling} (CASP), \citet{tully2015two,muller2016testing} ($CenA_{P1}$, $CenA_{P2}$).}
\label{figure:dimensions_all}
\end{figure}

The rms dimensions discussed in Sect. \ref{struc_dim} were calculated using only the Hough voters in each field and therefore removing the rest of the dwarfs in each field from consideration. For completeness we report the dimensions of the flattened structures proposed in this work considering the full dwarf sample in each field. The distributions for the rms length, rms height, and rms axis ratio are shown in Figs. \ref{figure:dim1all}, \ref{figure:dim2all}, and \ref{figure:dim3all}, respectively. Due to the inclusion of all dwarfs in the field, the rms height distribution extends much further toward higher numbers with a less pronounced peak. Consequently, the rms axis ratio follows this trend and shifts its peak toward higher values.

\end{document}